\documentclass[reprint,prb,twocolumn,superscriptaddress,aps,longbibliography]{revtex4-2}

\usepackage[colorlinks=true,citecolor=blue]{hyperref}%
\usepackage{graphicx}
\graphicspath{{./}{SM/figures/}}
\usepackage{bm}
\usepackage{amsmath}
\usepackage{amssymb}
\usepackage{comment}
\usepackage{braket}
\usepackage{orcidlink} 
\usepackage{physics}
\usepackage{balance}
\usepackage{placeins}
\usepackage[normalem]{ulem}
\usepackage{cancel}

\DeclareGraphicsExtensions{.png,.jpg,.eps}
\usepackage{xcolor}

\begin{document}

\title{\textbf{Real-space spectral functions of three-dimensional billion-size topological non-Hermitian matter with tensor networks}} 

\author{Yitao Sun\,\orcidlink{0009-0002-9479-7147}}
\thanks{yitao.sun@aalto.fi}
 \affiliation{Department of Applied Physics, Aalto University, 02150 Espoo, Finland}

\author{Jose L. Lado\,\orcidlink{0000-0002-9916-1589}}
 \affiliation{Department of Applied Physics, Aalto University, 02150 Espoo, Finland}

\author{Guangze Chen\,\orcidlink{0000-0002-1956-2519}}
\thanks{guangze@chalmers.se}
\affiliation{Department of Microtechnology and Nanoscience, Chalmers University of Technology, 41296 Gothenburg, Sweden}

\date{\today}%

\begin{abstract}
Non-Hermitian systems host unconventional topological phenomena, while real-space spectral calculations in three dimensions remain computationally challenging. Here, we develop a tensor-network framework that combines a quantics-based matrix-product-operator representation of large real-space tight-binding Hamiltonians with a non-Hermitian kernel polynomial method. 
The Hermitized spectral problem is retained within the compact tensor-network representation, enabling real-space spectral function calculations for systems exceeding one billion lattice sites.
Using this approach, we investigate a three-dimensional non-Hermitian higher-order topological insulator and resolve its corner-mode spectral response on a billion-site lattice, demonstrating the scalability and computational reach of the method.
By tuning the gain--loss strength, we identify distinct in-gap corner-mode features in the weak- and strong-gain--loss regimes.
Our results establish a scalable tensor-network framework for real-space spectral calculations in exceptionally large non-Hermitian lattice systems.
 
\end{abstract}

\maketitle

\section{Introduction}
 \label{intro}
Non-Hermitian systems~\cite{Ashida2021,Bergholtz2021RMP} provide a versatile framework for describing open quantum systems with gain--loss physics~\cite{PhysRevLett.123.073601,PhysRevB.110.155144,PhysRevResearch.6.023004,2025arXiv250905163P,breuer2002theory}, quantum transport~\cite{Helbig2019NHSE,Shen2024-og,PhysRevB.110.045138,PhysRevResearch.3.013208}, and topological phenomena~\cite{Bergholtz2021RMP,yao2018edge,lee2018anatomy,gong2018topological,okuma2020topological,kunst2018biorthogonal,Xiao2020}. 
Such systems have been realized in a variety of experimental platforms, including photonic and acoustic systems~\cite{Zhang2025,Zhang2021,Xiao2020,Gao2024,Hashemi2025}. Their spectra and eigenstates exhibit unique non-Hermitian phenomena, including exceptional points~\cite{Bergholtz2021RMP} in the spectrum and the non-Hermitian skin effect~\cite{yao2018edge}, which manifests in both spectral and spatial properties. These phenomena have no Hermitian counterparts. However, the primary numerical challenge stems from the complex spectra and large condition numbers of non-Hermitian operators, which often render large-scale simulations unstable and computationally expensive~\cite{RevModPhys.93.015008}. 
The difficulty is further intensified in three-dimensional systems, where the Hilbert space grows rapidly, and boundary effects become increasingly intricate.
\begin{figure}[t!]
    \centering
    \includegraphics[width=1\linewidth]{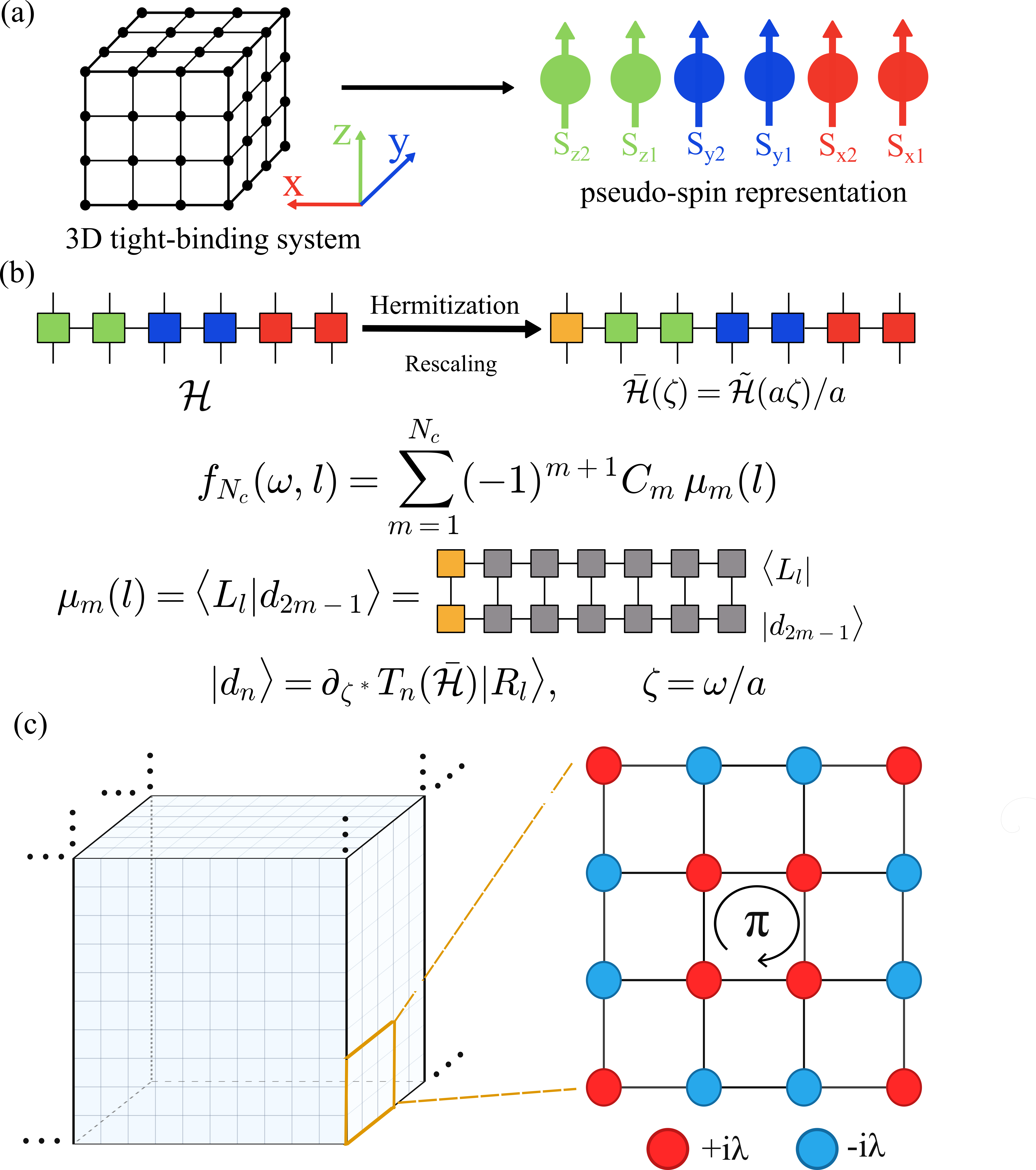}
    \caption{Schematic of the tensor network algorithm and the system of interest. Panel (a) shows an example of the tensorization of a 3D tight-binding system. Panel (b) shows MPO Hermitization and rescaling and the evaluation of spectral moments through MPS overlaps, with $C_m$ incorporating normalization and Jackson damping. Panel (c) illustrates the three-dimensional Benalcazar--Bernevig--Hughes (BBH)-type lattice models in which we demonstrate our methodology. The highlighted corner region is enlarged to show the non-Hermitian potential pattern. A $\pi$ flux is threaded through each plaquette via the Peierls substitution. Red and blue sites denote positive and negative imaginary onsite potentials, respectively. }
    \label{fig:algorithm}
\end{figure}

Tensor-network methods have proven highly effective for simulating quantum many-body systems by exploiting their low-entanglement structure, enabling controlled approximations while keeping computational costs manageable~\cite{PhysRevLett.69.2863,Schollwck2005,SciPostPhysLectNotes.133}. 
These techniques have been extended to non-Hermitian quantum many-body systems~\cite{PhysRevLett.130.100401,PhysRevResearch.6.043182,PhysRevLett.134.116605,Zhong2025,gzpj-pyj5}, offering a stable and scalable framework for computing spectral and dynamical properties.
The recent development of Quantics Tensor Cross Interpolation (QTCI) has further broadened the applicability of tensor-network techniques beyond the realm of quantum many-body systems~\cite{Oseledets2010,Oseledets2011,NunezFernandez2025learning,PhysRevLett.132.056501}.
QTCI provides an efficient way to compress high-dimensional data or operators across diverse research areas including quantum chemistry~\cite{PhysRevB.111.245115}, dynamical systems~\cite{Gourianov2025,Gourianov2022,Peddinti2024,niedermeierGP} and quantum many-body physics~\cite{PhysRevX.13.021015, PhysRevB.107.245135,PhysRevX.12.041018,10.21468/SciPostPhys.18.1.007,PhysRevB.110.035124,PhysRevResearch.7.023087}. 
Among those, one promising direction is to employ QTCI to compress large-scale real-space tight-binding Hamiltonians into compact tensor network representations.
For Hermitian systems, explorations have been conducted in mean-field simulations, real-space topology, momentum-space spectral function calculation and excitonic excitation in large-scale super-moir\'e and quasi-crystalline systems~\cite{ Sun2025,Anto2026,9btt-y8sh,https://doi.org/10.48550/arxiv.2603.02011}.
For single-particle non-Hermitian Hamiltonians, however, there is no a priori
guarantee that the Hermitized operator and the propagated Chebyshev vectors
remain compressible in the quantics representation, so the applicability of
this approach to non-Hermitian spectral calculations has remained an open
question.

In this work, we develop a tensor-network methodology that combines a quantics-based matrix-product-operator (MPO) representation of large real-space tight-binding Hamiltonians with the non-Hermitian kernel polynomial method~\cite{PhysRevLett.130.100401}.
Within this framework, the Hermitized Hamiltonian is constructed directly in the MPO representation, while the Chebyshev recursion is carried out on matrix-product-state (MPS) vectors. This combination retains a compact tensor-network description throughout the calculation and enables real-space spectral calculations for three-dimensional non-Hermitian systems containing more than one billion lattice sites.
As an example, we apply the framework to a three-dimensional non-Hermitian higher-order topological insulator (HOTI) and resolve its corner spectral features on a lattice containing more than one billion sites. 
This large-scale calculation serves as a demonstration of the scalability and computational reach of the method. 
More generally, the framework extends quantics-based real-space tensor-network methods to non-Hermitian spectral calculations and provides a route to studying structured non-Hermitian lattice systems at exceptionally large scales. 
The paper is organized as follows. In Sec.~\ref{sec:method}, we give an overview of the methodology, including the construction of the MPO for non-Hermitian Hamiltonians and the non-Hermitian kernel polynomial method.
In Sec.~\ref{sec:result}, we first introduce the model and its exact-diagonalization reference results on a moderate-size lattice, and then present the calculations on billion-site systems together with a discussion of the applicability and limitations of the algorithm. We conclude in Sec.~\ref{sec:conclusion}. The appendices contain the benchmarks against exact diagonalization and sparse-matrix KPM, the convergence tests, the implementation details, and further large-scale and spatially varying examples.

\section{Tensorization and NHKPM for non-Hermitian tight-binding Hamiltonians}
\label{sec:method}
Here we summarize the formalism for compressing a tight-binding Hamiltonian as a tensor network; further details can be found in Refs.~\cite{Anto2026, Sun2025,NunezFernandez2025learning,PhysRevLett.132.056501}. 
Consider a non-interacting real-space tight-binding Hamiltonian
$
H = \sum_{\alpha\beta} t_{\alpha\beta}\, c^\dagger_{\alpha} c_{\beta}, 
$
where $\alpha$ and $\beta$ label lattice sites. For a system with $N=2^L$ sites, each lattice index can be uniquely encoded in binary form as
$\alpha=(s_1,s_2,\ldots,s_L)$, with $s_i\in\{0,1\}$. This binary encoding defines a pseudo-spin representation, in which the single-particle Hilbert space is mapped onto the Hilbert space of an $L$-site spin-$\frac{1}{2}$ system. In this basis, the Hamiltonian matrix elements
$ 
H_{\alpha\beta}
= H_{(s_1,s_2,\ldots,s_L),(s'_1,s'_2,\ldots,s'_L)}
$ 
can be interpreted as those of an effective many-body Hamiltonian acting on a chain of $L$ pseudo-spins as shown in Fig.~\ref{fig:algorithm}(a).
Within this representation, the Hamiltonian admits an MPO form
\begin{equation}
\mathcal{H}_{(s_1,s_2,\ldots,s_L),(s'_1,s'_2,\ldots,s'_L)}
=
\Gamma^{(1)}_{s_1,s'_1}
\Gamma^{(2)}_{s_2,s'_2}
\cdots
\Gamma^{(L)}_{s_L,s'_L},
\end{equation}
where each tensor $\Gamma^{(n)}$ carries two physical indices $(s_n,s'_n)$ associated with the local two-dimensional pseudo-spin space and two virtual indices that are contracted with neighboring tensors. The dimension $D_H$ of the virtual indices defines the MPO bond dimension, which controls the expressive power and computational complexity of the representation.  
 
For simplicity, we next restrict our discussion to the MPO form of an onsite potential and first-neighbor hopping for a one-dimensional system~\cite{Sun2025,Anto2026}. 
The onsite modulations are introduced by allowing the diagonal terms to be purely imaginary, $t_{\alpha\alpha}= i\mathcal{U}_{\alpha}$, where the spatial profile $\mathcal{U}_{\alpha}$ can be compressed with the QTCI algorithm~\cite{Sun2025,PhysRevLett.132.056501}.
Here, the quantics representation specifies how the real-space index is
factorized into binary degrees of freedom, while QTCI provides a numerical
procedure for constructing the corresponding low-rank tensor
representation. Given a spatially smooth function, QTCI adaptively samples selected values of the
function and reconstructs a tensor-train representation through cross
interpolation, without evaluating the function on all lattice sites~\cite{PhysRevLett.132.056501,NunezFernandez2025learning}.
The resulting tensor can then be converted into the diagonal MPO entering
the Hamiltonian.
The accuracy of this construction is controlled by the
QTCI tolerance, while the required bond dimension depends on the
compressibility of the spatial profile.
The hopping term can be represented using compact shift MPOs
$\mathcal{H}_{\rm up}$ and $\mathcal{H}_{\rm down}$, which implement
forward and backward nearest-neighbor translations,
$\mathcal{H}_{\rm up}|x\rangle=|x+1\rangle$ and
$\mathcal{H}_{\rm down}|x\rangle=|x-1\rangle$, respectively, subject
to the chosen boundary conditions.
For uniform amplitudes, the hopping term can therefore be expressed as
$
\mathcal{H}_{\mathrm{hopping}} = t_{+} \mathcal{H}_{\mathrm{up}} + t_{-}\mathcal{H}_{\mathrm{down}},
$
where $t_{+}$ and $t_{-}$ can be different.
Position-dependent amplitudes, such as the dimerized and sign-staggered
hoppings are included by multiplying the shift MPOs
with diagonal MPOs of the hopping amplitudes, constructed with QTCI in the
same way as the onsite term.
Together with the diagonal onsite MPO, this yields a compact tensor-network representation of non-Hermitian tight-binding Hamiltonians with spatially varying nearest-neighbor hoppings and onsite terms.

Our methodology leverages the non-Hermitian kernel polynomial method
(NHKPM)~\cite{PhysRevLett.130.100401}. For a diagonalizable
non-Hermitian Hamiltonian $H$, the right and left eigenvectors satisfy
\begin{equation}
H|\Psi_{R,n}\rangle = E_n|\Psi_{R,n}\rangle,
\qquad
\langle\Psi_{L,n}|H = E_n\langle\Psi_{L,n}|,
\end{equation}
with the biorthogonal normalization
$\langle\Psi_{L,m}|\Psi_{R,n}\rangle=\delta_{mn}$.
The corresponding two-dimensional spectral operator and the spectral
function associated with a pair of probe states are
\begin{equation}
\begin{aligned}
\delta^2(\omega-H)
&=\sum_n\delta^2(\omega-E_n)
|\Psi_{R,n}\rangle\langle\Psi_{L,n}|,\\
f_{\phi\psi}(\omega)
&=\langle\phi|\delta^2(\omega-H)|\psi\rangle,
\end{aligned}
\label{eq:nh_spectral_function}
\end{equation}
where $|\phi\rangle$ and $|\psi\rangle$ are arbitrary probe states.
NHKPM evaluates this quantity by introducing the enlarged Hermitian
Hamiltonian and probe states
\begin{equation}
\begin{aligned}
\tilde H(\omega)
&=
\begin{pmatrix}
0 & \omega I-H \\
\omega^*I-H^\dagger & 0
\end{pmatrix},\\[3pt]
|L_\phi\rangle&=
\begin{pmatrix}0\\|\phi\rangle\end{pmatrix},
\qquad
|R_\psi\rangle=
\begin{pmatrix}|\psi\rangle\\0\end{pmatrix},
\end{aligned}
\label{eq:hermitized_hamiltonian}
\end{equation}
so that
\begin{equation}
f_{\phi\psi}(\omega)
=
\frac{1}{\pi}
\partial_{\omega^*}
\langle L_\phi|\tilde H^{-1}(\omega)|R_\psi\rangle .
\label{eq:nhkpm_green}
\end{equation}
For the kernel polynomial method (KPM) calculation, a fixed
positive scale $a$, including a small safety margin, is chosen such that
the spectrum of the rescaled Hermitized Hamiltonian remains inside the
Chebyshev interval $(-1,1)$ throughout the calculation. Introducing the
dimensionless energy $\zeta=\omega/a$, we write
\begin{equation}
\bar H(\zeta)
=
\frac{\tilde H(a\zeta)}{a}.
\label{eq:nhkpm_scaled_hamiltonian}
\end{equation}
Because the Hermitized spectrum is symmetric about zero, no additional
spectral shift is required. In the auxiliary basis
$|\uparrow\rangle=(1,0)^T$ and $|\downarrow\rangle=(0,1)^T$, we define
$\sigma^+=|\uparrow\rangle\langle\downarrow|$ and
$\sigma^-=|\downarrow\rangle\langle\uparrow|$. The derivative insertion
with respect to the scaled energy is then
\begin{equation}
\mathcal D_{\zeta}
\equiv
\partial_{\zeta^*}\bar H
=
\sigma^-\otimes I
=
\begin{pmatrix}0&0\\I&0\end{pmatrix}.
\label{eq:nhkpm_derivative_insertion}
\end{equation}
Defining the spectral density in the scaled complex-energy plane as
$F_{\phi\psi}(\zeta)=a^2 f_{\phi\psi}(a\zeta)$, the KPM expansion takes
the form
\begin{equation}
\begin{aligned}
F_{\phi\psi}(\zeta)
=\frac{2}{\pi}\sum_{m=1}^{\infty}(-1)^{m+1}
\langle L_\phi|
\partial_{\zeta^*}T_{2m-1}(\bar H)
|R_\psi\rangle .
\end{aligned}
\label{eq:nhkpm_chebyshev}
\end{equation}
The derivative is evaluated recursively following Ref.~\cite{PhysRevLett.130.100401}. In practical calculations, the expansion
is truncated and regularized with a Jackson kernel~\cite{RevModPhys.78.275,Jackson1912}.
For the site-resolved spectral function we choose
$|\phi\rangle=|\psi\rangle=|l\rangle$ and denote the result by
$f(\omega,l)$, with total spectral density
$\rho(\omega)=\sum_l f(\omega,l)$.

We now implement this construction in the quantics tensor-network
representation. Writing the non-Hermitian Hamiltonian as an $L$-site MPO
$\mathcal H$, Hermitization is achieved by adding one auxiliary two-level
site to the MPO chain,
\begin{equation}
\tilde{\mathcal H}(\omega)
=
\sigma^{+}\otimes(\omega\mathcal I-\mathcal H)
+
\sigma^{-}\otimes(\omega^*\mathcal I-\mathcal H^\dagger),
\label{Hermitized_MPO}
\end{equation}
where $\mathcal I$ is the identity MPO. The Hermitized MPO therefore has
$L+1$ tensor sites, and the auxiliary site is placed at the largest
quantics scale to keep the MPO bond dimension controlled. The rescaled
MPO $\bar{\mathcal H}(\zeta)=\tilde{\mathcal H}(a\zeta)/a$ is used in
the Chebyshev recursion below.
Rather than explicitly constructing the polynomial operators
$T_n(\bar{\mathcal H})$, the numerical calculation propagates with MPS
vectors. For a site-localized probe
$|R_l\rangle=|\uparrow\rangle\otimes|l\rangle$, the ordinary Chebyshev
vectors, their initial values, and their recurrence are
\begin{equation}
\begin{aligned}
|t_n\rangle &= T_n(\bar{\mathcal H})|R_l\rangle,\\
|t_0\rangle &= |R_l\rangle,\\
|t_1\rangle &= \bar{\mathcal H}|R_l\rangle,\\
|t_{n+1}\rangle &=2\bar{\mathcal H}|t_n\rangle-|t_{n-1}\rangle.
\end{aligned}
\label{eq:nhkpm_mps_recurrence}
\end{equation}
Here the vectors are defined for $n=0,1,2,\ldots$, and the recurrence
applies for $n\geq1$. In parallel, the derivative vectors obey
\begin{equation}
\begin{aligned}
|d_n\rangle &=\partial_{\zeta^*}T_n(\bar{\mathcal H})|R_l\rangle,\\
|d_0\rangle &=0,\\
|d_1\rangle &=\mathcal D_{\zeta}|R_l\rangle,\\
|d_{n+1}\rangle
&=2\mathcal D_{\zeta}|t_n\rangle
+2\bar{\mathcal H}|d_n\rangle
-|d_{n-1}\rangle.
\end{aligned}
\label{eq:nhkpm_derivative_recurrence}
\end{equation}
The recurrence is propagated through all polynomial orders up to
$2N_c-1$, while only the odd-order derivative moments enter the
spectral expansion.
For the $m$th retained odd contribution, with $m=1,\ldots,N_c$ and
polynomial order $2m-1$, we evaluate a scalar overlap with
$|L_l\rangle=|\downarrow\rangle\otimes|l\rangle$.
The moments, the kernel-weighted spectral function in the scaled plane,
and its conversion to the original energy plane are
\begin{equation}
\begin{aligned}
\mu_m(l)&\equiv\langle L_l|d_{2m-1}\rangle,\\
F_{N_c}(\zeta,l)
&=
\frac{2}{\pi}
\sum_{m=1}^{N_c}
(-1)^{m+1}
g_{2m-2}\,\mu_m(l),\\
f_{N_c}(\omega,l)&=a^{-2}F_{N_c}(\omega/a,l),
\end{aligned}
\label{eq:LDOS_MPS}
\end{equation}
where $g_k$, with $k=0,\ldots,2N_c-1$, are the Jackson damping factors
for an expansion truncated at maximum order $2N_c-1$.
The contribution of order $2m-1$ is weighted by $g_{2m-2}$, i.e., the
damping factor is indexed by the polynomial degree of
$\partial_{\zeta^*}T_{2m-1}(\bar{\mathcal H})$, as in the public
implementation of Ref.~\cite{PhysRevLett.130.100401}. Indexing by
$2m-1$ instead changes the peak amplitudes by $\lesssim0.5\%$ at
$N_c=500$ and leaves the peak positions and integrated spectral weights
unchanged.
At each retained odd order, the scalar overlap $\mu_m(l)$ is evaluated
before the final kernel-weighted summation, so no MPS representing the
complete Chebyshev sum needs to be constructed. The implementation used
for Figs.~\ref{fig:3} and \ref{fig:4} is specified in
Appendix~\ref{sec:sm_resources}.

During the recurrence, MPO--MPS applications and MPS linear combinations
are compressed using both a maximum retained MPS bond dimension $\chi$
and a compression cutoff. The MPS bond dimension $\chi$ is distinct from
the Hamiltonian MPO bond dimension $D_H$, whose accuracy is controlled by
the QTCI construction tolerance. Detailed convergence tests with respect
to $\chi$, the MPS compression cutoff, and the Chebyshev expansion order
$N_c$ are presented in Appendix~\ref{sec:sm_convergence}. The Hamiltonian MPO
representation is validated separately through full-matrix reconstruction
(Appendix~\ref{sec:sm_resources}).
\section{Results}
\label{sec:result}
\subsection{Benchmark in a moderate-size system}
\begin{figure}[t!]
    \centering
    \includegraphics[width=\linewidth]{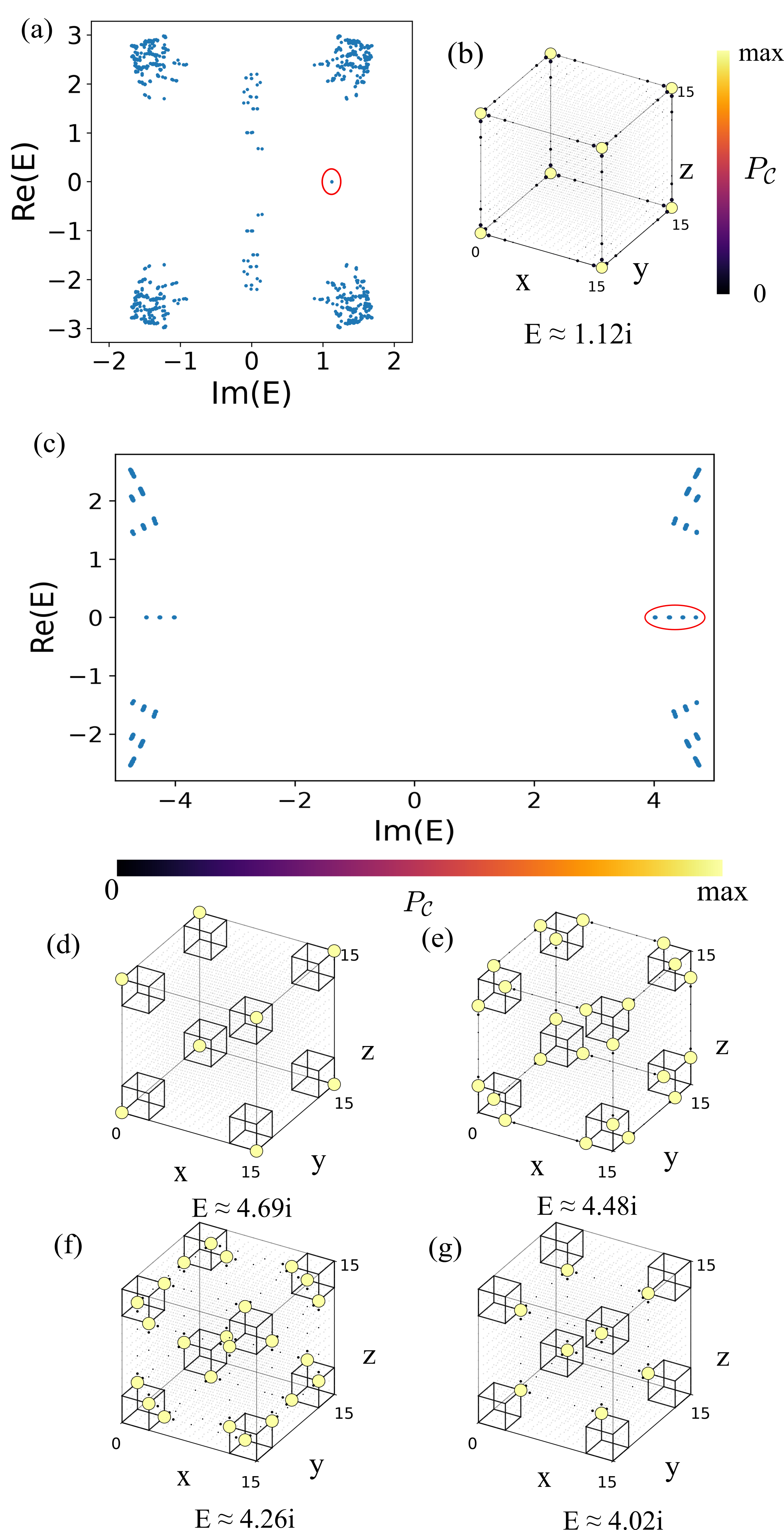}
    \caption{Exact-diagonalization results for the open $16^3$ lattice. Panels (a) and (b) correspond to the weak gain--loss regime ($\lambda=2$), while panels (c)--(g) show the strong gain--loss regime ($\lambda=5$). The red circles indicate the spectral regions containing the selected corner-mode subspaces. Panels (b) and (d)--(g) show their biorthogonal spatial weights $P_{\mathcal C}(s)$ without energy broadening. The four strong-gain--loss subspaces contain 8, 24, 24, and 8 states and exhibit 1, 3, 3, and 1 dominant vertices per outer corner cube, respectively. All lattice sites are retained. In panels (d)--(g), the boxes outline the eight outer $4^3$ corner cubes. The energy annotations give the mean eigenvalue of each selected subspace, and the color scale runs from zero to its spatial maximum. Site coordinates are zero based. }
    \label{fig:2}
\end{figure}

To demonstrate the methodology, we consider the Hamiltonian
$H=H_0+H_V$ corresponding to Fig.~\ref{fig:algorithm}(c).
We denote the lattice sites by
$\mathbf r=(r_x,r_y,r_z)$, with
$r_\mu\in\{1,\ldots,\ell_\mu\}$, where $\ell_\mu$ is the physical side length along direction $\mu=x,y,z$.
The Hermitian part is
\begin{align}
H_0 = \sum_{\mathbf r}\sum_{\mu=x,y,z}
\Big[
s_\mu(\mathbf r)\,t_\mu(\mathbf r)\,
c^\dagger_{\mathbf r+\hat\mu}c_{\mathbf r}
+\mathrm{h.c.}
\Big],
\label{eq:H0}
\end{align}
where terms for which $\mathbf r+\hat\mu$ lies outside the lattice
are omitted, corresponding to open boundary conditions in all three
directions.
The staggered sign structure is
\begin{align}
s_x(\mathbf r)&=1,\nonumber\\
s_y(\mathbf r)&=(-1)^{r_x},\\
s_z(\mathbf r)&=(-1)^{r_x+r_y},\nonumber
\end{align}
and the hopping amplitudes are dimerized according to
\begin{equation}
t_\mu(\mathbf r)=
\begin{cases}
t_1,& r_\mu\ \mathrm{odd},\\
t_2,& r_\mu\ \mathrm{even}.
\end{cases}
\label{eq:t_uniform}
\end{equation}
For the even linear system sizes considered here, the outermost bond
adjacent to each surface has amplitude $t_1$.
The staggered signs generate a $\pi$ flux through each plaquette.
Together with the dimerized hoppings, this realizes the
three-dimensional Benalcazar--Bernevig--Hughes (BBH)-type lattice~\cite{Benalcazar2017Science,Benalcazar2017PRB} considered below.

For the spatially modulated gain--loss potential, the zero-based
coordinates, four-site block indices, and modulation used in both the
exact-diagonalization and tensor-network calculations are defined by
\begin{equation}
\begin{aligned}
n_\mu&=r_\mu-1,\qquad
b_\mu=\left\lfloor\frac{n_\mu}{4}\right\rfloor,\\
s(u)&=
\sqrt{2}\cos\!\left(
\frac{\pi u}{2}+\frac{\pi}{4}
\right),\\
V(\mathbf r)
&=
\prod_{\mu=x,y,z}
s(n_\mu)s(b_\mu).
\end{aligned}
\label{eq:onsite_V}
\end{equation}
The factor $s(n_\mu)$ produces a period-four modulation at the site
level, while $s(b_\mu)$ modulates successive four-site blocks.
Their product therefore gives a period-16 pattern along each spatial
direction.
The corresponding non-Hermitian onsite term is
\begin{equation}
H_V
=
i\lambda
\sum_{\mathbf r}
V(\mathbf r)\,
c^\dagger_{\mathbf r}c_{\mathbf r}.
\label{eq:gain_loss}
\end{equation}
\begin{figure}[t!]
    \centering
    \includegraphics[width=0.95\linewidth]{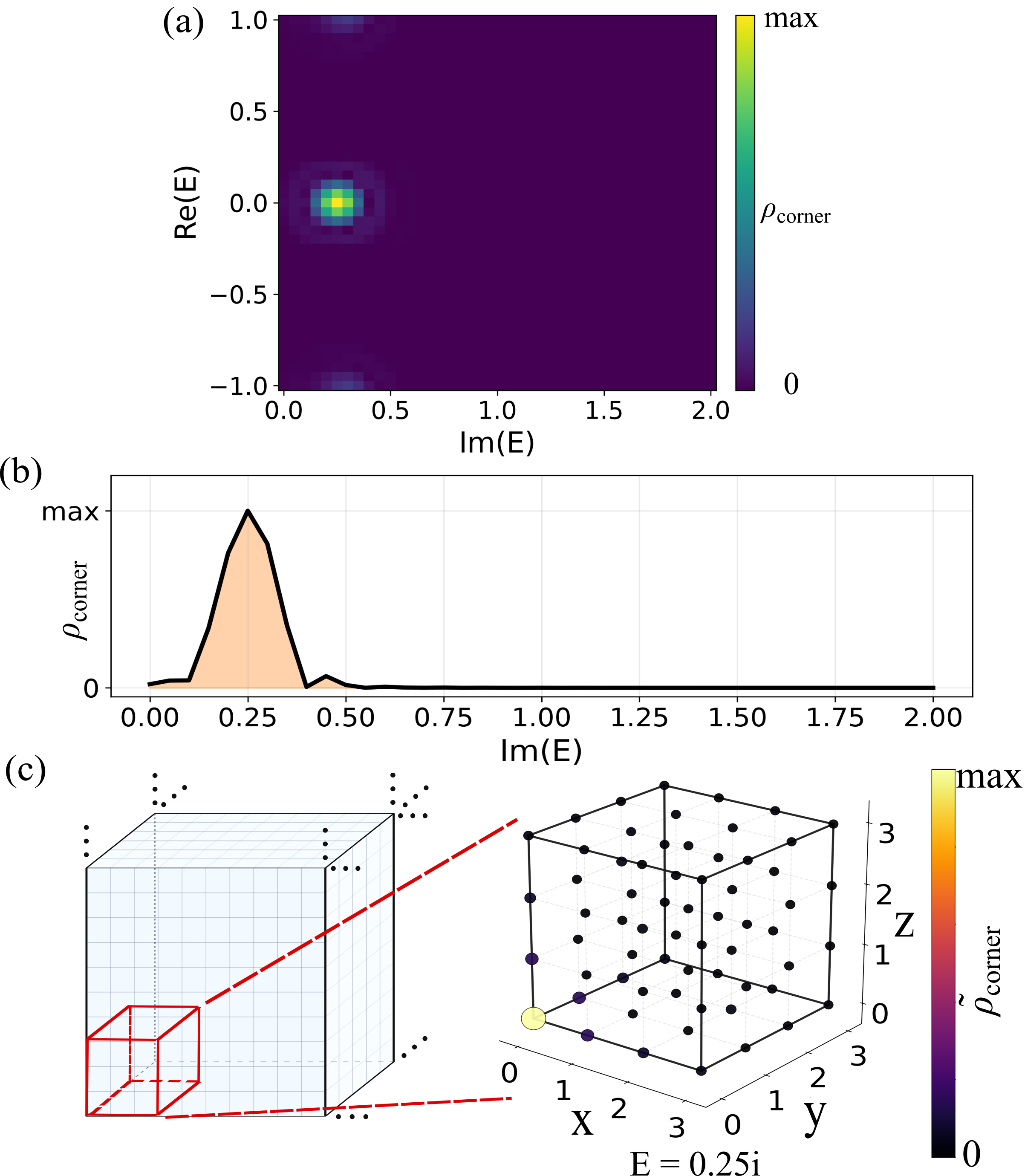}
    \caption{Tensor-network results for the weak gain--loss regime ($\lambda = 1.4$) in the $1024^3\approx 10^9$ system. Panel (a) shows the corner local density of states (LDOS) on the complex plane with a single broadened in-gap mode separated from the bulk continuum,
    while (b) shows a cut of (a) with $\mathrm{Re}(E) = 0$.
    Panel (c) shows the spatial LDOS of the in-gap mode within one representative corner cube, demonstrating its corner localization. The cube sites are labeled by the zero-based coordinates $n_\mu=r_\mu-1$. Panel (a) uses $\operatorname{Re}(E)\in[0,1]$ and $\operatorname{Im}(E)\in[0,2]$ with spacing $0.05$, giving 861 independently sampled energies. The negative-real-energy half is obtained by symmetry reflection. Panel (b) contains the 41-point $\operatorname{Re}(E)=0$ cut. Panel (c) samples all 64 cube sites at $E=0.25i$. The color scales run from zero to max. In (c), max denotes $1$ for the normalized profile $\widetilde\rho_{\rm corner}$. }
    \label{fig:3}
\end{figure}
The hopping parameters are chosen as $t_1 = 1$ and $t_2 = 1.4$, such that the Hermitian part $H_0$
lies in the higher-order topological regime of the three-dimensional
BBH-type model.
The role of the gain--loss term is then to reshape the complex spectrum and
spectrally separate the corner-sector eigenvalues from the bulk
continuum. 
As a consequence, the corner modes become isolated by a
finite line gap in the complex energy plane, facilitating their direct
identification through real-space spectral functions.

We first consider a system of size $16^3$, which can be solved by exact diagonalization (ED). As shown in Fig.~\ref{fig:2}, two representative gain--loss amplitudes are examined.
For the ED spatial maps in Fig.~\ref{fig:2}, we resolve the corner-mode
subspaces directly, without energy broadening. We plot
$P_{\mathcal C}(s)=|\sum_{n\in\mathcal C}w_n(s)|$,
where $w_n(s)=\langle s|\Psi_{R,n}\rangle\langle\Psi_{L,n}|s\rangle$
and the left and right eigenvectors are biorthonormal.
Complete degenerate or nearly degenerate multiplets are retained.
Within each indicated spectral region, outer-corner multiplets are identified
by requiring more than half of the absolute diagonal-projector weight to lie
in the eight outer $4^3$ corner cubes. For $\lambda=5$, the selected multiplets
have more than $90\%$ of this weight in those cubes, whereas the other
nearby multiplets have less than $0.2\%$.
The modulus is taken after summing the selected states, and the resulting
profile is shown on the full lattice, including its spatial tails.
Each outer corner cube corresponds to the region shown in the tensor-network
spatial maps below. The Gaussian-broadened ED spectral functions used for
the quantitative benchmarks in Appendix~\ref{sec:sm_benchmark} are evaluated
separately, including all eigenmodes.
The site labels use the zero-based
coordinates $n_\mu=r_\mu-1$.
For weaker gain--loss strength $\lambda=2$, the spectrum of the $16^3$ lattice contains a single isolated corner-mode sector, whose spatial weight remains localized at the outer corners of the entire system as shown in Fig.~\ref{fig:2}(b).
Increasing the gain--loss strength to $\lambda=5$ qualitatively modifies the complex-energy spectrum, leading to four spectrally separated families of in-gap corner modes for $\mathrm{Im}(E)>0$.
The corresponding spatial-weight patterns, shown in Fig.~\ref{fig:2}(d)--(g), reveal a hierarchical splitting of the corner-mode manifold. The mode with the largest imaginary energy remains concentrated near the exterior corner of the corner cubes, whereas modes at lower energies become progressively localized at their inner corners. 

These ED results provide reference mode positions and spatial profiles
for a direct quantitative validation of the tensor-network calculation.
We therefore perform matched ED, sparse-KPM, and tensor-network (TN) KPM calculations
on the same open $16^3$ lattice and at identical parameters. For
$\lambda=2$, TN and sparse KPM with $N_c=200$ agree over the sampled
complex-energy window with a relative complex-data error of
$\sim10^{-5}$. The corresponding spatial profiles, computed with
$N_c=500$, also agree closely. The TN--sparse errors over the three
coordinate edges are below $10^{-5}$ at both $\lambda=2$ and $5$.
After normalization over the same set of sites, the TN--ED profile
differences are $\sim10^{-3}$ at both gain--loss strengths. The larger residual difference in
the ED comparison also includes the distinct Gaussian and finite-order
KPM broadening procedures. Detailed benchmarks and convergence tests
are provided in Appendices~\ref{sec:sm_benchmark} and \ref{sec:sm_convergence}.

\begin{figure}[t!]
    \centering
    \includegraphics[width=0.95\linewidth]{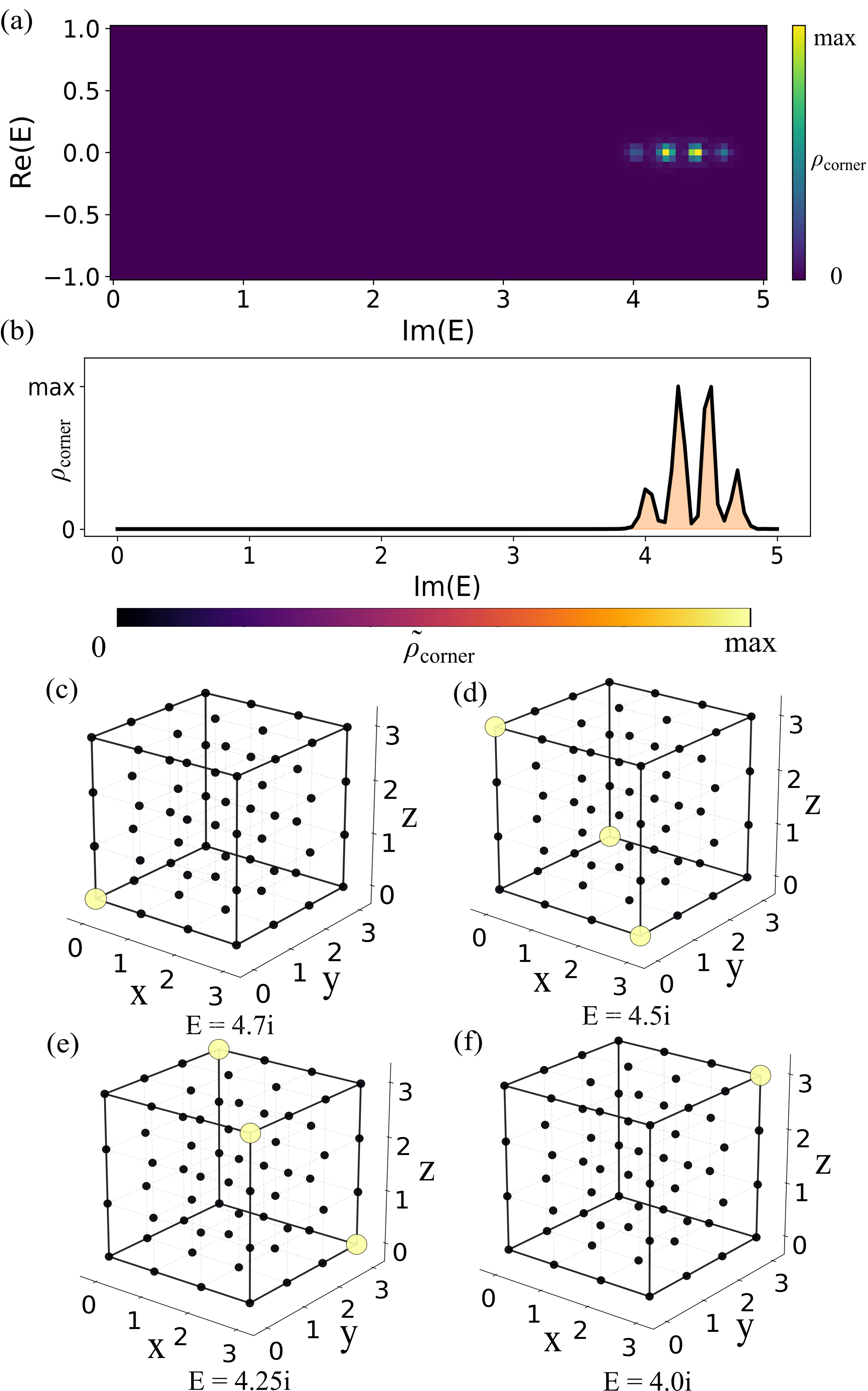}
    \caption{Tensor-network results for the strong gain--loss regime ($\lambda=5$) in the $1024^3\approx 10^9$ system.
    Panel (a) shows the corner LDOS on the complex plane where multiple isolated modes emerge along the imaginary-energy axis. The spectral cut (b) at $\mathrm{Re}(E) = 0$ further probes the hierarchical splitting of the corner modes. Panels (c)--(f) show the corner LDOS for the in-gap modes at different complex energies, revealing distinct localization patterns within the same corner cube as in Fig.~\ref{fig:3}. Panel (d) was computed with a larger bond cap $\chi=120$ as an additional accuracy margin, since the four in-gap modes lie close in energy. The grid in (a) is $\operatorname{Re}(E)\in[0,1]$, $\operatorname{Im}(E)\in[0,5]$, with spacing $0.05$ and 2121 energy coordinates per vertex class. The negative-real-energy half is obtained by symmetry reflection. The line cut in (b) contains 101 energies from $0$ to $5.00i$, with spacing $0.05$. Panels (c)--(f) each sample 64 cube sites at $E=4.70i,4.50i,4.25i,4.00i$, respectively. The color scales run from zero to max. In (c)--(f), max denotes $1$ for the normalized profiles $\widetilde\rho_{\rm corner}$.}
    \label{fig:4}
\end{figure}

\subsection{Topological corner modes in billion-site systems} 
We now turn to systems of size $1024^3 \approx 10^9$, far beyond the
reach of exact diagonalization. Unless stated otherwise, the
tensor-network calculations use $N_c=500$ retained odd Chebyshev orders,
corresponding to a maximum polynomial order $2N_c-1=999$, a maximum
propagated MPS bond dimension $\chi=100$, and the spectral rescaling
factor $a=12$.
We use an MPS compression cutoff of $10^{-8}$ (operation-specific settings are given in Appendix~\ref{sec:sm_resources}), while the QTCI construction tolerance of the Hamiltonian MPO is
controlled independently and is also set to $10^{-8}$. For the 
model considered here, the Hamiltonian MPO has maximum bond dimension
$D_H=5$, increasing to $6$ after Hermitization. For the cubic lattices,
the coordinate bits are arranged in $z/y/x$ blocks, with $x$ the
fastest real-space coordinate. Detailed convergence, Hamiltonian-MPO
reconstruction, and resource benchmarks are provided in
Appendices~\ref{sec:sm_convergence}--\ref{sec:sm_capabilities}.

Because the corner sectors are symmetry related, we focus on one
representative corner region. To characterize the corner spectrum, we
average the site-resolved spectral response over a selected set of corner
probe sites $\mathcal P$,
\begin{equation}
\rho_{\rm corner}(\omega)
=
\frac{1}{|\mathcal P|}
\sum_{s\in\mathcal P} |f(\omega,s)|.
\end{equation}
The probe set $\mathcal P$ is specified for each calculation. For Fig.~\ref{fig:3}(a), $\mathcal P$ contains a single corner probe
site, whereas for Fig.~\ref{fig:4}(a) it contains the eight vertices of the
outermost $4\times4\times4$ corner cube.
At the energies of the resolved in-gap features, we further examine the
spatial local density of states (LDOS) distribution over a larger displayed corner region
$\mathcal R$, the 64 sites of the outermost $4\times4\times4$ corner cube, which we label in the figures by the zero-based coordinates $n_\mu=r_\mu-1=0,\ldots,3$. For these spatial profiles, we use
\begin{equation}
\widetilde{\rho}_{\rm corner}(s;\omega)=
\frac{|f(\omega,s)|}
{\max_{s'\in\mathcal R}|f(\omega,s')|},
\qquad s\in\mathcal R .
\end{equation}
Thus, $\rho_{\rm corner}(\omega)$ characterizes the corner spectrum,
whereas $\widetilde{\rho}_{\rm corner}(s;\omega)$ shows the normalized
spatial distribution of the local spectral response.

As shown in Fig.~\ref{fig:3}(a), a broadened but well-resolved corner mode is observed near $\mathrm{Im}(E) = 0.25$ for the system with the weaker gain--loss strength of $\lambda = 1.4$.
We use $\lambda=1.4$ rather than the value $\lambda=2$ of Fig.~\ref{fig:2}(a,b),
because the in-gap corner mode is then more clearly separated from the bulk
continuum. At $\lambda=2$, the isolated corner-mode sector of the $16^3$
lattice is partly a finite-size feature and need not remain spectrally
isolated in much larger systems.
The corresponding LDOS in Fig.~\ref{fig:3}(c) remains localized at the outer part of the corner cube, demonstrating the persistence of the corner-mode sector in the billion-site calculation.
For the stronger gain--loss strength shown in Fig.~\ref{fig:4}(a), the corner-mode sector splits into four distinct in-gap modes.
The associated LDOS patterns in Fig.~\ref{fig:4}(c)--(f) reproduce the same hierarchical localization sequence observed in the ED calculations, ranging from modes localized at the outermost corner of the corner cube to modes concentrated at its innermost corner. 
The strong gain--loss results show a qualitatively similar hierarchy of corner features to the $16^3$ ED calculation, while the weak gain--loss calculation exhibits the same single corner-sector response. These results show that the tensor-network framework resolves the spectral response of the non-Hermitian HOTI at exceptionally large system sizes.
 
Beyond the periodic model considered above, the framework can also
accommodate spatially varying Hamiltonian coefficients without requiring
a model-specific analytic MPO representation. As an explicit test, in
Appendix~\ref{sec:sm_domainwall} we consider a smooth gain--loss domain wall whose
onsite potential is supplied only through pointwise evaluations and
tensorized directly using QTCI. The resulting TN-KPM spectra agree with
independent sparse-KPM calculations over the complete sampled
complex-energy grid with relative errors of order $10^{-4}$.
This benchmark demonstrates that the construction is not restricted to
the periodic onsite pattern of Eq.~\eqref{eq:gain_loss} and provides a practical
route toward more general nonperiodic and incommensurate real-space
Hamiltonians.
More broadly, the same strategy can be applied to structured
non-Hermitian tight-binding models such as quasiperiodic
systems~\cite{PhysRevB.104.024201,PhysRevLett.129.113601,
PhysRevLett.122.237601,PhysRevLett.132.263801},
fractals~\cite{Manna2023}, and moir\'e systems with large real-space
unit cells~\cite{PhysRevB.111.085120,dl59-vl7v,dfzm-hj41,Yuan2026}.
The computational cost is controlled primarily by the Hamiltonian MPO
bond dimension $D_H$, the retained MPS bond dimension $\chi$, and the
Chebyshev expansion order $N_c$, while the spectral resolution in the
complex-energy plane is set by the finite polynomial order.

Sparse-matrix Chebyshev methods provide a complementary route to
large-scale real-space spectral calculations and have reached systems
containing tens of billions of orbitals in Hermitian
settings~\cite{Joao2020KITE}. The present tensor-network formulation is
not intended to replace such approaches, but instead exploits a different
source of scalability. In the quantics representation, a system of
$N=2^L$ sites is encoded by a tensor network of length
$L=\log_2 N$, so that, provided the Hamiltonian MPO and propagated MPS
vectors remain compressible, increasing the physical system size does not
require storing objects whose dimensions grow explicitly with $N$.
We demonstrate this scaling by extending the calculation from
$1024^3=2^{30}$ to
$16384^3=2^{42}$ physical sites, which adds
only 12 quantics sites before Hermitization. At the largest size, we
evaluate a full 11,211-point complex-energy window and resolve the
spatial corner response at its sampled spectral maximum. The recurrence
has a measured peak process resident set size (RSS) of $1.92$--$1.96$ GiB and requires
approximately 67 minutes per sampled energy point on one CPU thread,
for the three energies $E=0.20i,0.25i,0.30i$ at the stated truncation
settings. These timings include first-call compilation within the
recurrence and exclude Hamiltonian construction and Hermitization. These results illustrate that the practical range of
the method is governed primarily by the compressibility and bond dimensions
of the Hamiltonian MPO and propagated MPS vectors, rather than by the
physical system size alone. Detailed convergence and resource-scaling
benchmarks are provided in Appendices~\ref{sec:sm_convergence} and \ref{sec:sm_capabilities}.

\section{Conclusion}
\label{sec:conclusion}

In summary, we have established a tensor-network framework for computing
real-space spectral functions of large non-Hermitian tight-binding
systems by combining quantics-based MPO representations with a
non-Hermitian kernel polynomial method, while retaining the Hermitized
spectral problem and Chebyshev recursion within compressed tensor-network
representations. As a representative application, we applied the
framework to a three-dimensional non-Hermitian higher-order topological
insulator with over one billion sites and resolved distinct corner-mode spectral responses in the
weak- and strong-gain--loss regimes. In particular, increasing the
gain--loss strength leads from a single isolated corner-mode sector to a
hierarchically split set of in-gap corner features with distinct spatial
localization patterns. The numerical reliability and broader computational
capability of the framework are further supported by quantitative
same-lattice benchmarks against sparse-matrix KPM and exact diagonalization,
systematic convergence tests, and calculations extending to
$16384^3=2^{42}\simeq4.4\times10^{12}$ physical sites.
In addition, a
smooth spatially varying gain--loss potential constructed directly from
pointwise evaluations with QTCI demonstrates that the approach is not
restricted to the periodic onsite pattern considered in Sec.~\ref{sec:result}.
More broadly, these results establish a scalable and flexible framework
for real-space spectral studies of structured non-Hermitian systems,
complementary to large-scale sparse-matrix approaches and extending
tensor-network electronic-structure methods far beyond the system sizes
accessible to direct diagonalization.
 
\textbf{Acknowledgment}---We acknowledge the computational resources provided by the Aalto Science-IT project
and the financial support from InstituteQ, 
the
Research Council of Finland (project No. 370912), 
the Finnish Ministry
of Education and Culture through the Quantum Doctoral Education Pilot Program (QDOC VN/3137/2024-OKM-4), the
Finnish Quantum Flagship (project No. 358877, Aalto University),
the Finnish Centre of Excellence in Quantum Materials QMAT (No. 374166),
and the ERC Consolidator Grant ULTRATWISTROICS (Grant agreement no.
101170477). G.C.~is supported by the European Union's Horizon Europe programme HORIZON-MSCA-2023-PF-01-01 via the project 101146565 (SING-ATOM).
We thank A. Moustaj, T. Ant\~ao, and P. Shen for useful discussions. 
The code used for this work can be found in Ref.~\cite{myrepo}.

\appendix
\setlength{\textfloatsep}{12pt plus 2pt minus 2pt}
\setlength{\dbltextfloatsep}{12pt plus 2pt minus 2pt}
\setlength{\floatsep}{10pt plus 2pt minus 2pt}
\setlength{\intextsep}{12pt plus 2pt minus 2pt}

\section{Benchmark against exact diagonalization and sparse KPM}
\label{sec:sm_benchmark}

To validate the numerical implementation, we compare tensor-network (TN) KPM
with sparse-matrix KPM and exact diagonalization (ED) on an open $16^3$
lattice, using the Hamiltonian of Eqs.~\eqref{eq:H0}--\eqref{eq:gain_loss} with $t_1=1$ and $t_2=1.4$.
Throughout the appendices, sites are labeled by the zero-based
coordinates $n_\mu=r_\mu-1$, so that the corner is $(0,0,0)$
and each coordinate of the $16^3$ lattice runs from 0 to 15. The coordinate
$x$ of the domain wall below is the same zero-based coordinate.
The two KPM implementations use identical spectral rescaling and polynomial
order, while ED provides an independent reference for both the local spectral
distribution and its spatial profile.
Here $\rho\equiv f(\omega,s)$ denotes the site-resolved biorthogonal
spectral function at complex energy $\omega$ and probe site $s$, using
the definition of $f$ in Eq.~\eqref{eq:nh_spectral_function}. All spectral plots display
$|\rho|$, as the biorthogonal local spectral weights may
take either sign, while finite-order KPM together with MPS truncation can also
produce signed oscillations in regions of small spectral weight. When
quantifying the difference between TN and sparse KPM, however, we retain the
full complex spectral function. For two sets of sampled values $\rho_{A,i}$ and
$\rho_{B,i}$, where $i$ runs over the sampled energies or sites, we define
the relative error
\begin{equation}
\mathcal{E}(A,B)=
\frac{\sum_i|\rho_{A,i}-\rho_{B,i}|}
     {\sum_i\left(|\rho_{A,i}|+|\rho_{B,i}|\right)},
\label{eq:sm_complex_error}
\end{equation}
which is used throughout the appendices to quantify the difference
between two sampled spectra or spatial profiles. Since the denominator
contains both magnitudes, $\mathcal{E}$ is symmetric in $A$ and $B$, lies
between 0 and 1, and stays finite where one of the two sets is small.

\begin{figure}[!htb]
\centering
\includegraphics[width=\linewidth]{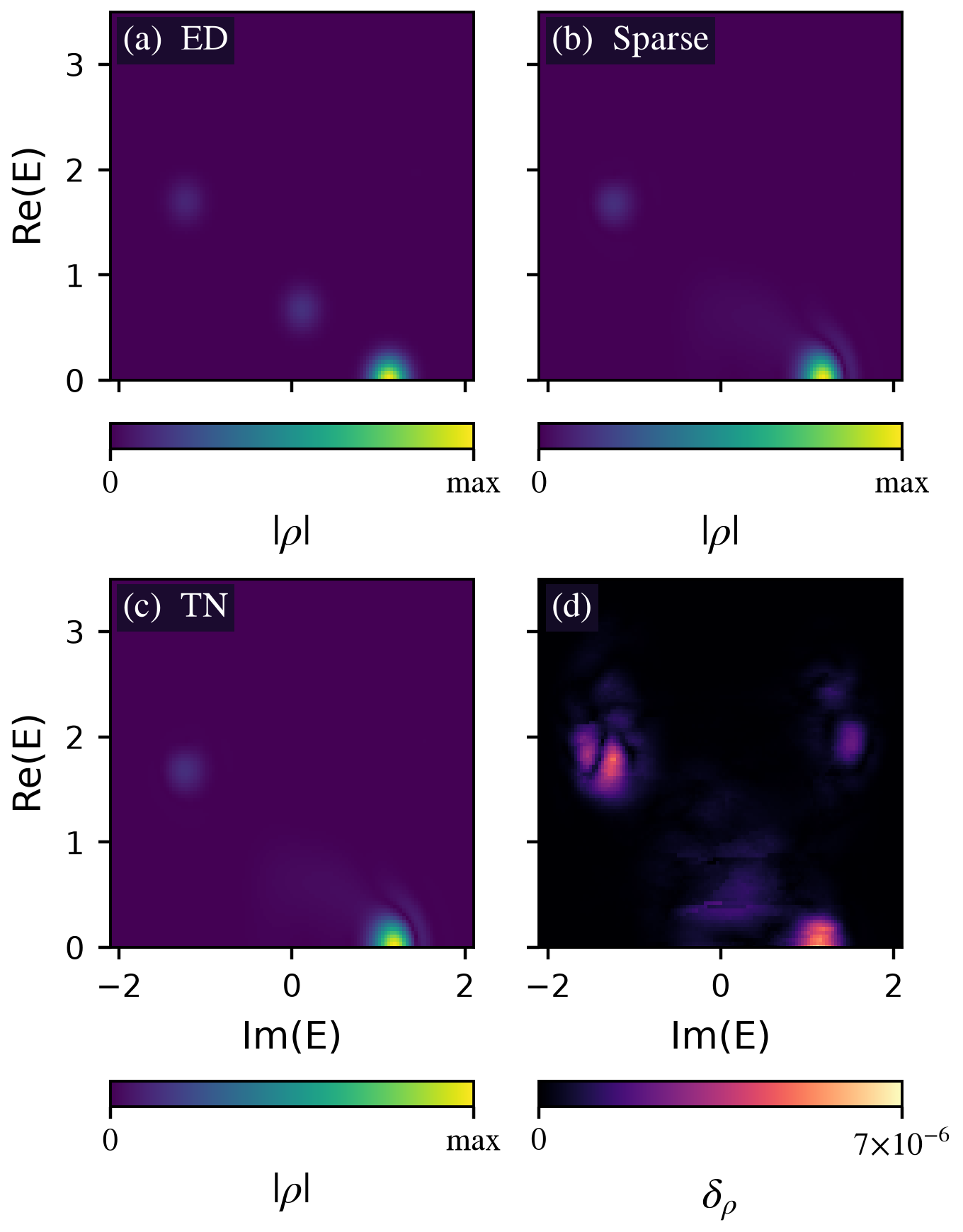}
\caption{Local spectrum at the corner $(0,0,0)$ of the open $16^3$ lattice with
$\lambda=2$. (a) ED with biorthogonal local weights and Gaussian kernel
$\exp[-|E-E_n|^2/\eta^2]/(\pi\eta^2)$, using $\eta\simeq0.2015$
to match the dominant KPM peak's FWHM.
(b,c) Sparse and TN KPM with $N_c=200$, $a=12$, and TN bond cap
$\chi=100$/compression cutoff $10^{-8}$. The energy grid
covers $\mathrm{Re}(E)\in[0,3.5]$ and $\mathrm{Im}(E)\in[-2.1,2.1]$
with step $0.035$ in both directions. Panels (a) to (c) use linear
color scales marked only by zero and their maximum. The two KPM maps share
one scale, while ED has a separate one. (d) Complex-data error
$\delta_\rho=|\rho_{\mathrm{TN}}-\rho_{\mathrm{sparse}}|/
\max_E|\rho_{\mathrm{sparse}}|$, displayed directly in scientific notation.
No map is normalized by its own peak.}
\label{fig:sm_periodic_spectra}
\end{figure}

\subsection{Site-resolved complex-energy spectrum}

Figure~\ref{fig:sm_periodic_spectra} compares the local spectrum at the
corner site $(0,0,0)$ for $\lambda=2$.
All three methods resolve the dominant corner response near the
imaginary-energy axis and the weaker features at finite real energy.
Sparse and TN KPM agree across the sampled complex-energy window.
Their relative complex error is
$\mathcal{E}(\mathrm{TN},\mathrm{sparse})\sim 10^{-5}$.
The ED map includes all eigenvalues with biorthogonal local weights,
broadened by a normalized two-dimensional Gaussian with
$\eta\simeq0.2015$. This broadening is introduced for comparison with
the finite-order KPM spectrum so that
its width matches the full width at half maximum (FWHM) of the dominant
sparse-KPM peak, approximately $0.336$. The Gaussian and finite-order
KPM kernels still differ in shape. On the common
$\operatorname{Re}(E)=0$ grid, the sampled ED and KPM maxima occur at
$1.12i$ and $1.19i$, respectively.
This benchmark is evaluated separately
from Fig.~\ref{fig:2}, whose ED spatial maps show selected corner-mode
subspace weights without energy broadening.
Taken together, these comparisons show that TN KPM reproduces
the sparse-KPM spectrum to high numerical accuracy, while remaining
consistent with the independently obtained ED spectral features despite
the different broadening kernels.

\begin{figure}[!htb]
\centering
\includegraphics[width=\linewidth]{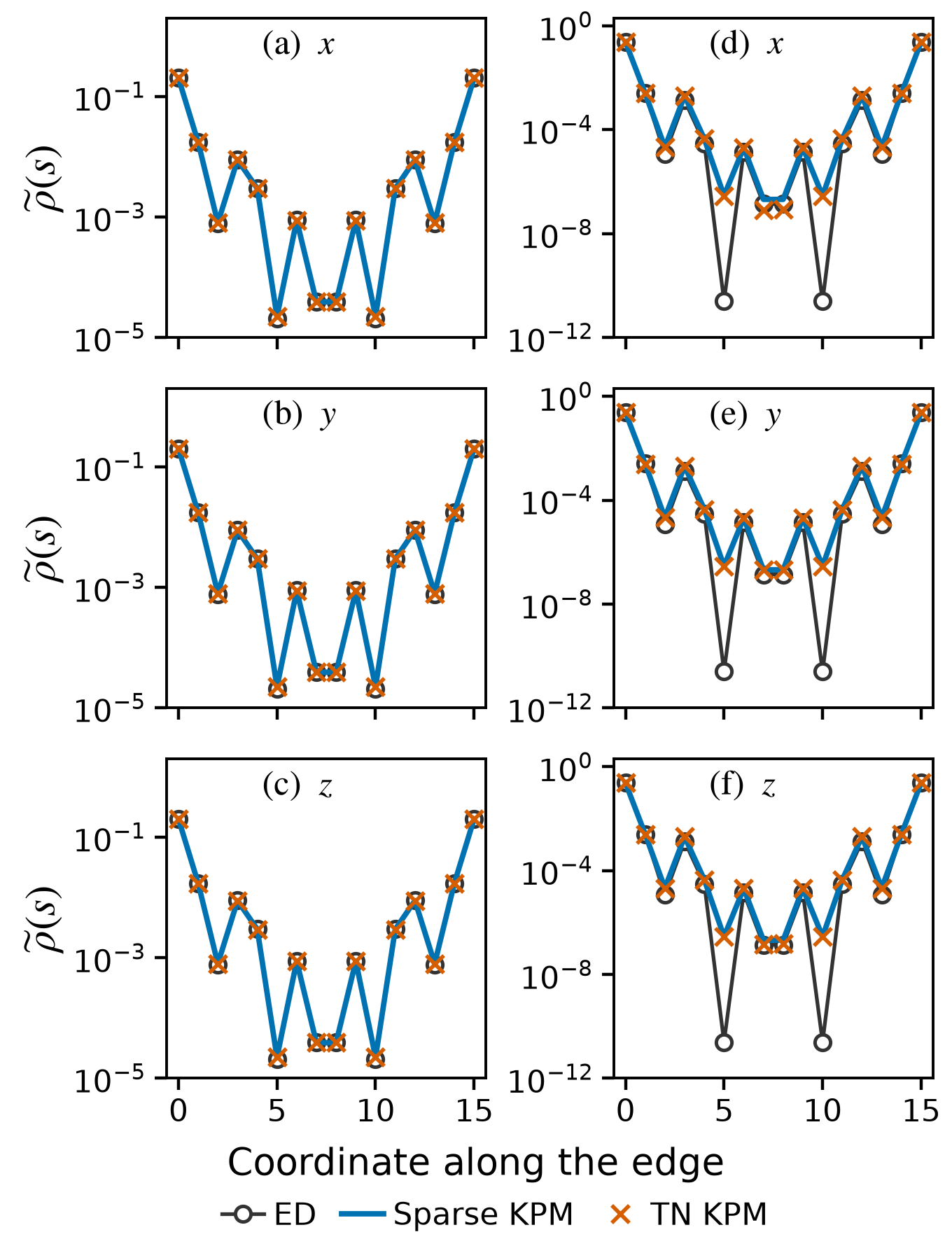}
\caption{Normalized spatial response $\widetilde\rho(s)$ on the open $16^3$
lattice. (a--c) The edges $(x,0,0)$, $(0,y,0)$ and $(0,0,z)$ at
$\lambda=2$, $E=1.12i$. (d--f) The same edges at $\lambda=5$,
$E=4.69i$. ED uses Gaussian broadening $\eta=0.05$. Both KPM methods
use $N_c=500$ and $a=12$, with TN $\chi=100$ and cutoff $10^{-8}$.
Each method is normalized over the same edge sites, as defined in Eq.~\eqref{eq:sm_normalized_profile},
with one normalization across all three edges of a given case.}
\label{fig:sm_spatial_profiles}
\end{figure}

\subsection{Corner-mode spatial profiles}
 
Figure~\ref{fig:sm_spatial_profiles} extends the benchmark to the spatial
structure of the response and, in particular, to the stronger gain--loss
regime. We compare the response along the three complete coordinate edges
at $E=1.12i$ for $\lambda=2$ and $E=4.69i$ for $\lambda=5$.
These edges contain 46 distinct sites, including the common origin once.
To compare the spatial distributions, we normalize each method over the
same set of sites $\mathcal S$.
\begin{equation}
\widetilde\rho(s)=
\frac{|\rho(s)|}{\sum_{s'\in\mathcal S}|\rho(s')|}.
\label{eq:sm_normalized_profile}
\end{equation}
The three methods reproduce the enhanced response at the outer corners
and the alternating spatial structure along each edge. For the normalized
profiles, $\mathcal{E}(\widetilde\rho_{\mathrm{TN}},
\widetilde\rho_{\mathrm{ED}})\sim10^{-3}$ at both gain--loss strengths.
Part of these deviations originates from the different spectral broadening
procedures. ED uses
Gaussian broadening, whereas KPM has the finite-order broadening associated
with the polynomial expansion. The ED comparison is therefore not expected
to exhibit the same level of pointwise agreement as a direct TN--sparse KPM
comparison.
Comparing TN and sparse KPM directly at identical KPM settings removes
the broadening difference associated with ED. The unnormalized local
magnitudes closely overlap at both gain--loss strengths, and the relative
errors over the 46 sites are below $10^{-5}$.
Taken together, these comparisons show that the TN calculation accurately
reproduces both the overall spatial structure and the local spectral weights
obtained from sparse KPM in the weak and strong gain--loss regimes,
while remaining consistent with the independently computed ED profiles.

\begin{figure}[!htb]
\centering
\includegraphics[width=\linewidth]{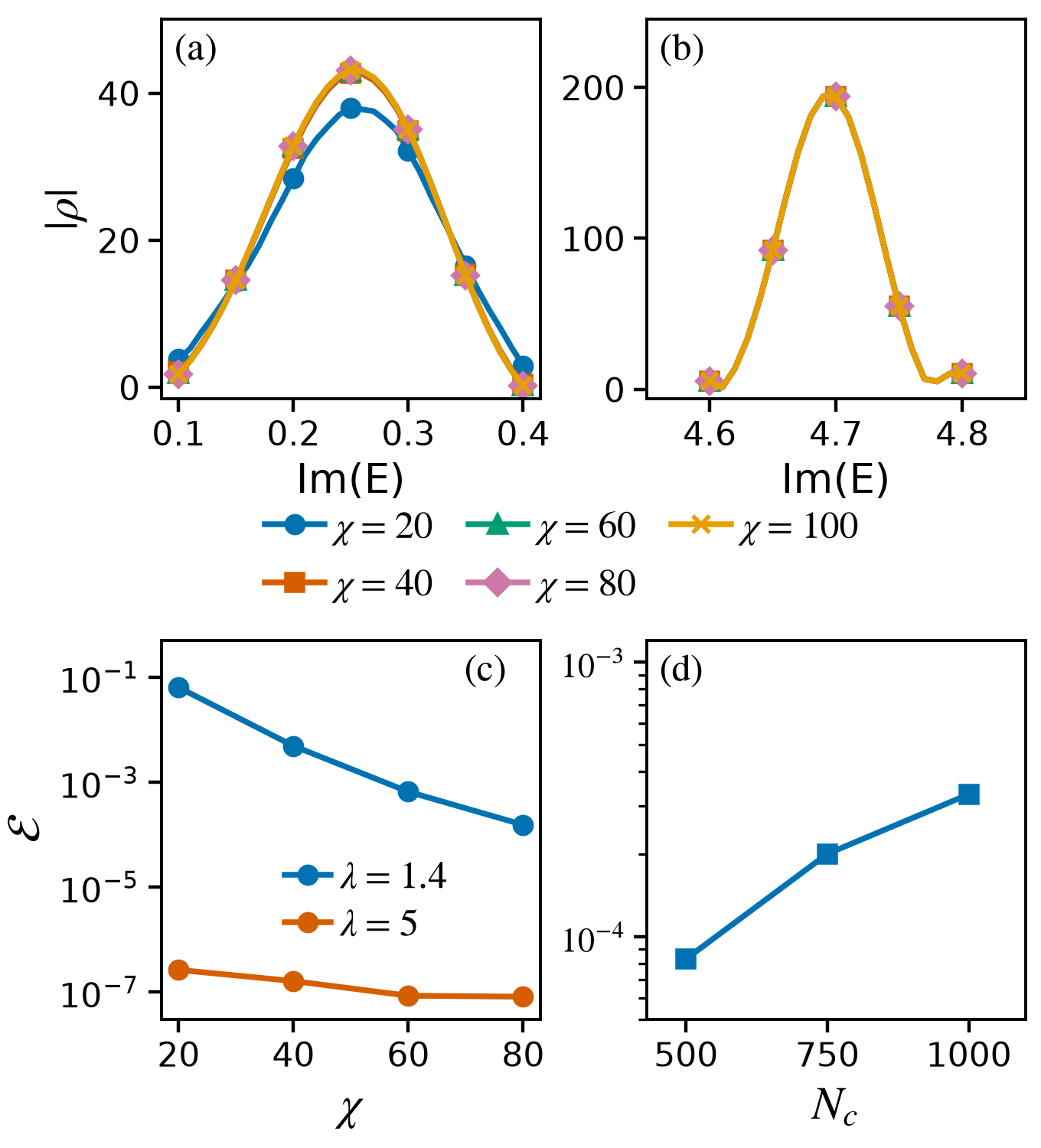}
\caption{MPS bond-dimension convergence on the $1024^3$ lattice at
$\mathrm{Re}(E)=0$. (a,b) Spectral cuts at $N_c=500$ for
$\lambda=1.4$ and $5$, respectively. (c) Relative complex errors
against $\chi=100$ over each sampled window. (d) Relative complex error
between $\chi=100$ and $150$ for $\lambda=1.4$. For each $N_c$, the
five energies $\mathrm{Im}(E)=0.23,0.24,0.25,0.26,0.27$ are combined
using Eq.~\eqref{eq:sm_complex_error} to give one error value.
All panels use $a=12$ and cutoff $10^{-8}$.}
\label{fig:sm_chi}
\end{figure}

\section{Convergence in bond dimension, polynomial order, and compression cutoff}
\label{sec:sm_convergence}

The convergence analysis uses the corner probe of an open
$1024^3$ lattice. We denote the maximum retained MPS bond dimension by
$\chi$ and the number of retained odd polynomial orders by $N_c$,
so that the largest retained order is $2N_c-1$.
Unless stated otherwise, the spectral rescaling factor is fixed at
$a=12$ and the MPS compression cutoff at $10^{-8}$.
We vary $\chi$ and $N_c$ separately to distinguish bond-dimension
truncation from finite-order spectral broadening. The dependence on the
compression cutoff is examined independently in
Fig.~\ref{fig:sm_cutoff} at fixed $\chi=100$ and $N_c=500$.

\subsection{MPS bond dimension}

Figure~\ref{fig:sm_chi}(a,b) shows the spectral cuts for
$\chi=20,40,60,80,100$ at fixed $N_c=500$.
In the weak gain--loss regime, the peak shape progressively approaches
the $\chi=100$ reference as the retained bond dimension increases.
The relative complex error over the sampled window decreases from
$\sim 10^{-1}$ at $\chi=20$ to $\sim 10^{-4}$ at
$\chi=80$. In the strong gain--loss regime, the curves around the
outermost peak already overlap closely, with corresponding relative
errors $\sim 10^{-7}$ or smaller.
To further show the adequacy of the $\chi=100$ reference, we perform
additional calculations with $\chi=150$ near the weak-regime peak.
For each $N_c=500,750,1000$, we combine the five common energies
using Eq.~\eqref{eq:sm_complex_error} to evaluate one relative complex
error between the $\chi=100$ and $150$ results. These three aggregate
errors are all $\sim 10^{-4}$ [Fig.~\ref{fig:sm_chi}(d)].
Thus, even as the polynomial expansion is extended to $N_c=1000$,
the residual dependence on the retained bond dimension remains at $\sim 10^{-4}$ over the sampled peak region. 
These results support $\chi=100$ as an adequate working bond dimension for the sampled spectral regions examined here.

\begin{figure}[!htb]
\centering
\includegraphics[width=\linewidth]{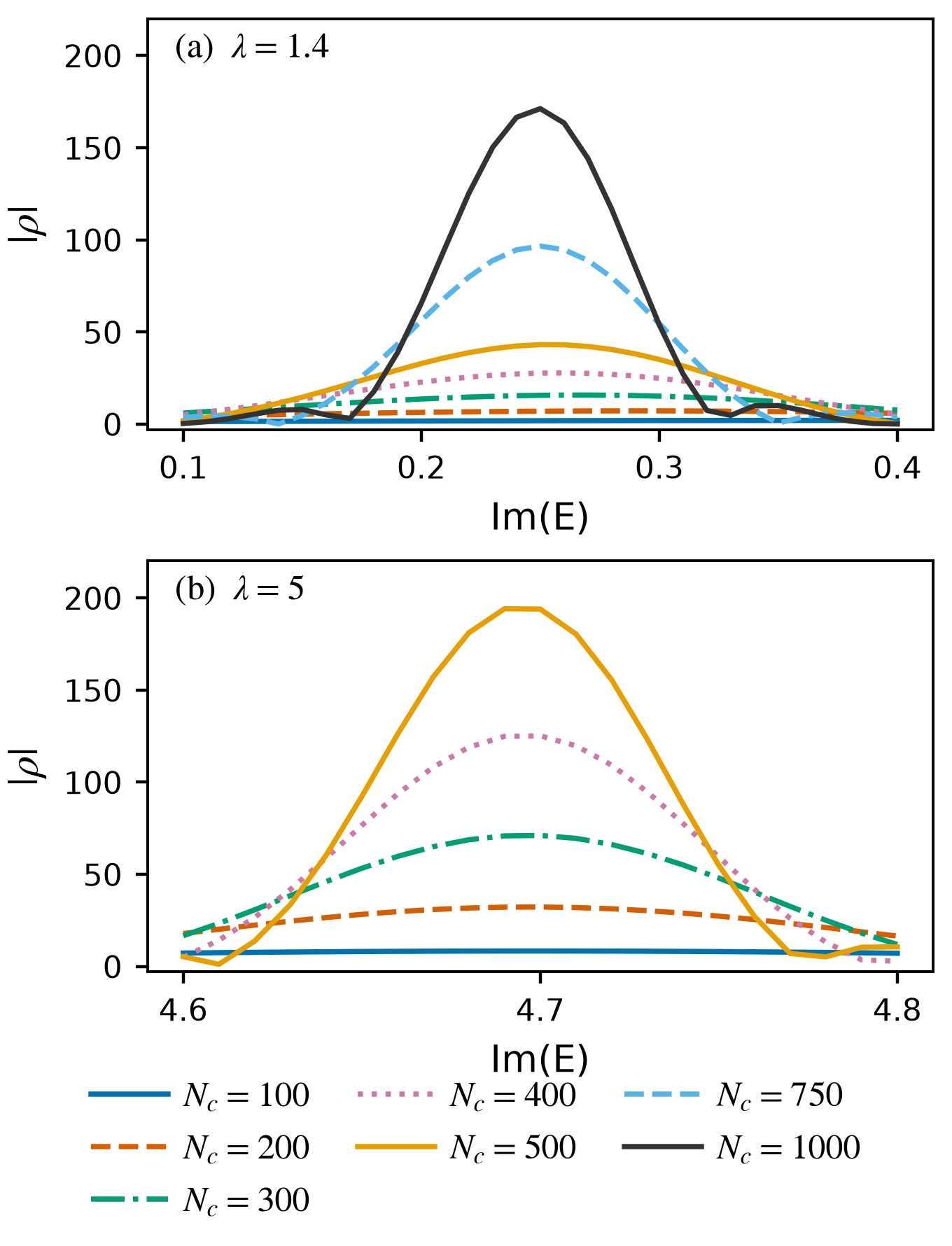}
\caption{Polynomial-order dependence on the $1024^3$ lattice for the corner
probe, $\mathrm{Re}(E)=0$, $\chi=100$, $a=12$ and cutoff $10^{-8}$.
(a) $\lambda=1.4$ with $N_c=100,200,300,400,500,750,1000$. The
$N_c\geq500$ curves use the extended-order calculations.
(b) $\lambda=5$ with $N_c=100$--$500$.
Both panels show physical $|\rho|$ sampled at spacing $0.01$.
FWHM values quoted in the text use linear interpolation at the
half-height crossings.}
\label{fig:sm_nc}
\end{figure}

\subsection{Polynomial order and spectral resolution}

Figure~\ref{fig:sm_nc} shows the dependence on the polynomial order
$N_c$ at fixed $\chi=100$. In the weak gain--loss regime, the sampled
peak position remains at $\mathrm{Im}(E)=0.25$ for
$N_c=500,750,1000$, while the FWHM decreases from $0.166$ to
$0.111$ and $0.083$, respectively. The peak height grows approximately
as $N_c^2$, so that the product of the peak height and the squared FWHM
stays nearly constant, consistent with a fixed integrated spectral weight. The peak narrowing is approximately
proportional to $1/N_c$, consistent with the
improved spectral resolution expected at higher polynomial order. The
strong gain--loss peak likewise becomes progressively sharper as $N_c$
is increased.
Overall, these results show that increasing
$N_c$ primarily improves the spectral resolution while leaving the
resolved peak position unchanged within the sampling accuracy.

\begin{figure}[!htb]
\centering
\includegraphics[width=\linewidth]{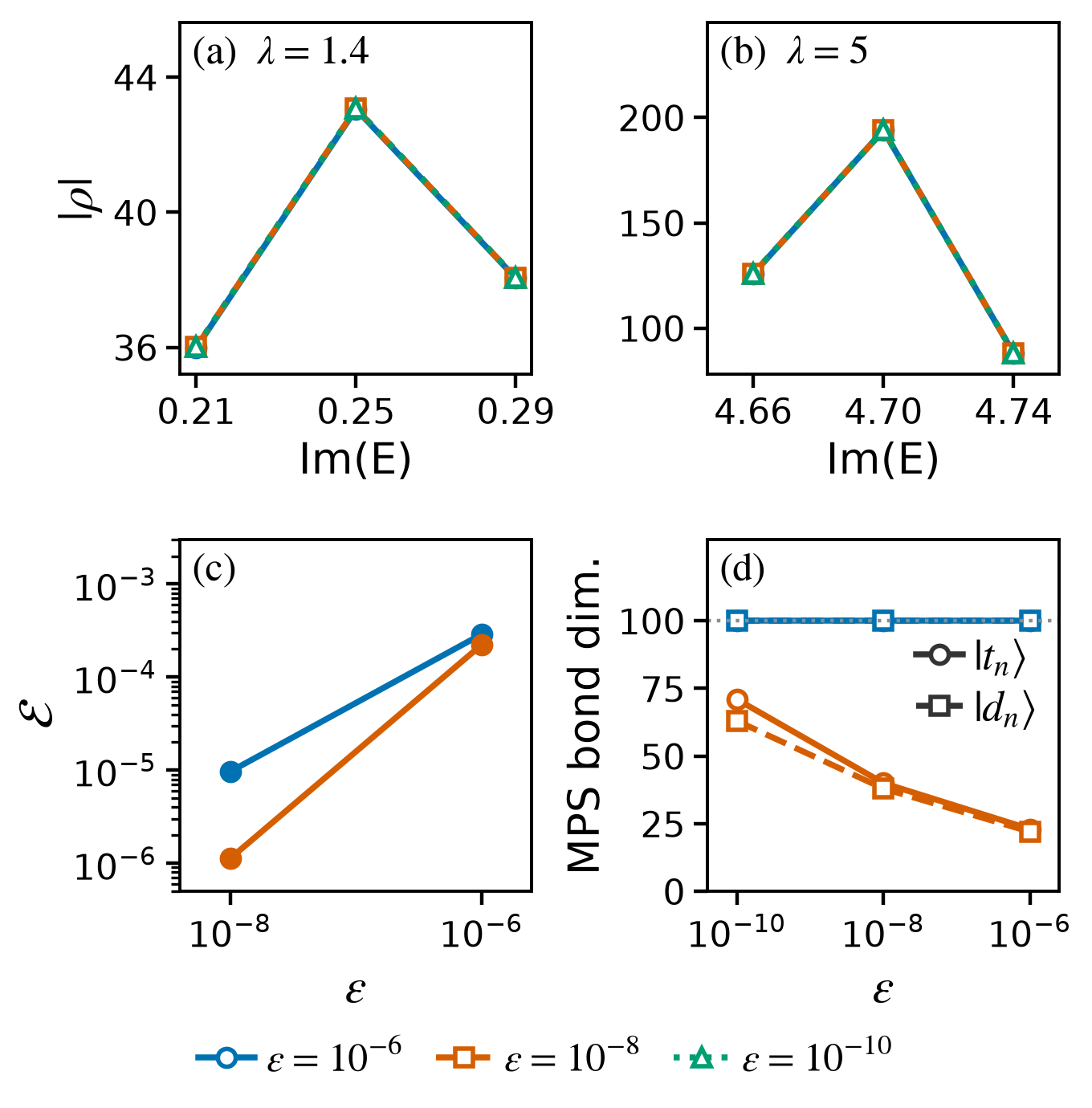}
\caption{MPS cutoff dependence for the corner probe on the open
$1024^3$ lattice, with $\chi=100$, $N_c=500$, $a=12$,
$t_1=1$, $t_2=1.4$, and fixed QTCI tolerance $10^{-8}$.
(a,b) Physical $|\rho|$ at the three selected energies for
$\lambda=1.4$ and $5$. Lines join these discrete samples only.
(c) Relative complex errors against cutoff $10^{-10}$, using
Eq.~\eqref{eq:sm_complex_error} on each three-energy set. The zero
reference error is omitted from the logarithmic axis.
(d) Largest MPS bond dimensions over the three energies and all
recorded odd orders, shown separately for $|t_n\rangle$ and
$|d_n\rangle$. In (c,d), blue and orange denote $\lambda=1.4$ and $5$,
respectively. The solid-circle and
dashed-square curves in (d) represent $|t_n\rangle$ and $|d_n\rangle$,
respectively. The dotted line marks $\chi=100$.}
\label{fig:sm_cutoff}
\end{figure}

\subsection{MPS compression cutoff}

\begin{figure}[!htb]
\centering
\includegraphics[width=\linewidth]{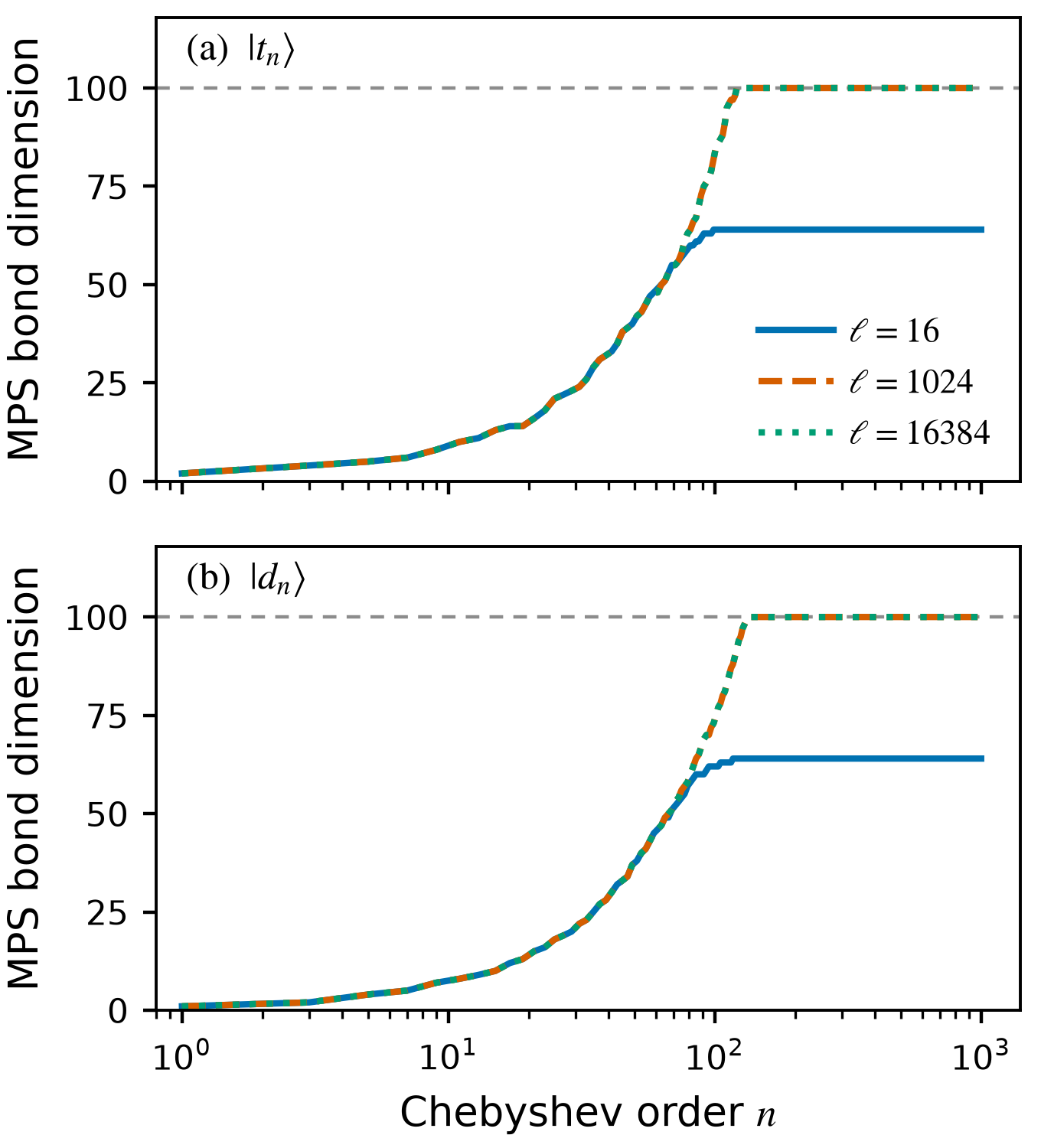}
\caption{Bond dimensions of (a) the ordinary Chebyshev MPS
$|t_n\rangle$ and (b) the derivative MPS $|d_n\rangle$ for side
lengths $\ell=16,1024,16384$. The largest dimension along each MPS is
recorded at odd polynomial orders through $n=999$. Calculations use
$E=0.25i$, $\lambda=1.4$, $N_c=500$, $a=12$ and $\chi=100$.
The dashed line marks the imposed cap. The horizontal axis is
logarithmic. Hamiltonian and Hermitized MPO dimensions are $5$ and $6$.}
\label{fig:sm_bond_growth}
\end{figure}

\begin{figure}[!htb]
\centering
\includegraphics[width=\linewidth]{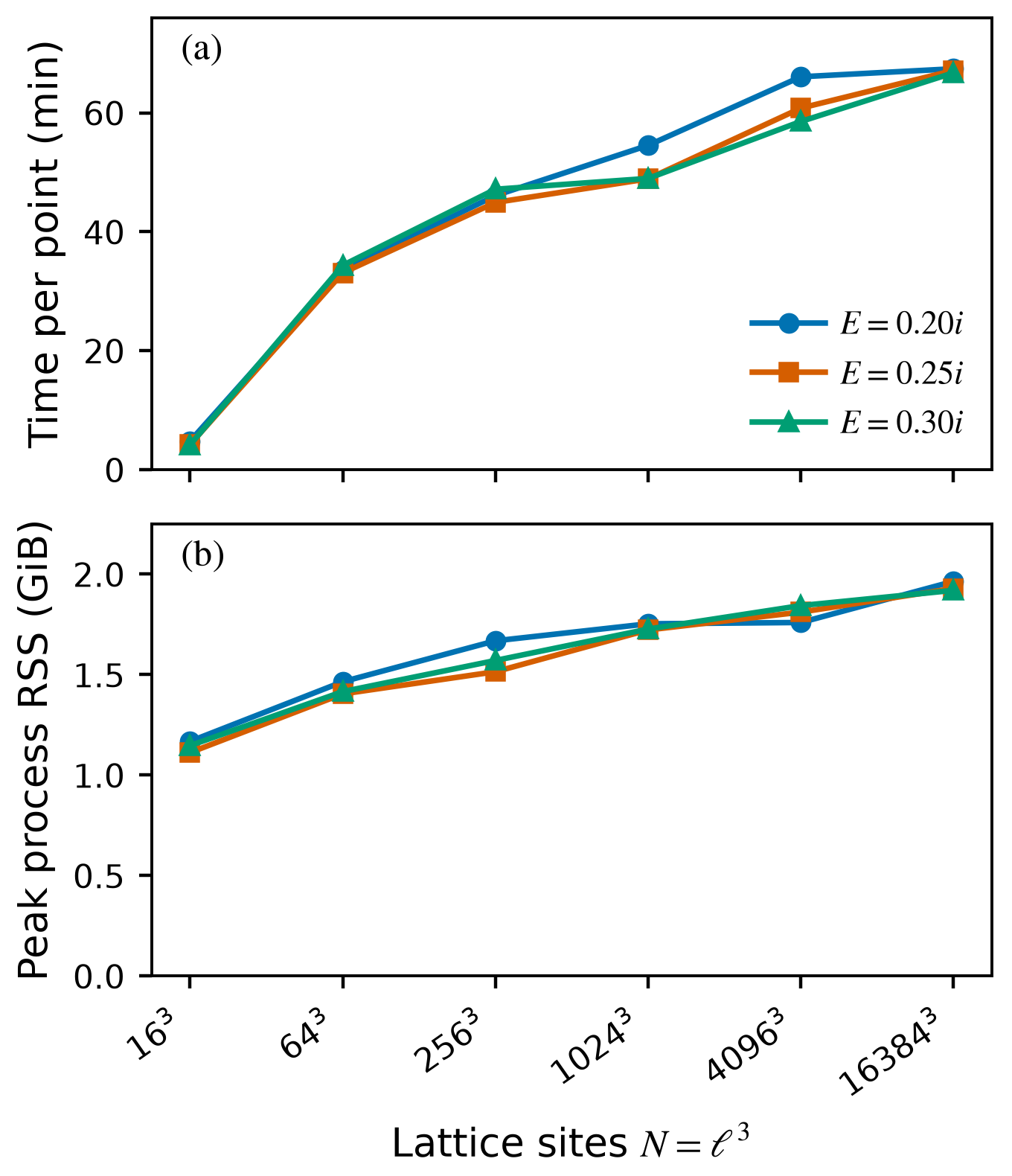}
\caption{Resource scaling for $E=0.20i,0.25i,0.30i$ and
$\ell=16,64,256,1024,4096,16384$. (a) KPM recurrence time per energy
point, including first-call compilation within the recurrence and excluding
Hamiltonian construction and Hermitization. (b) Measured peak resident
set size of the Julia process, including runtime and temporary allocations.
Parameters are $\lambda=1.4$, $N_c=500$, $\chi=100$, $a=12$,
cutoff $10^{-8}$ and corner probe $(0,0,0)$.
The horizontal axis shows
the physical site count $N=\ell^3$ on a logarithmic scale.}
\label{fig:sm_scaling}
\end{figure}

Figure~\ref{fig:sm_cutoff} examines the dependence on the MPS compression
cutoff $\varepsilon$ on the $1024^3$ lattice. We fix $\chi=100$,
$N_c=500$, $a=12$, the QTCI tolerance at $10^{-8}$, and use the corner
probe $(0,0,0)$. The tested compression cutoffs for MPS are
$\varepsilon=10^{-6},10^{-8},10^{-10}$. For $\lambda=1.4$, we sample
$E=0.21i,0.25i,0.29i$, while for $\lambda=5$ we sample
$E=4.66i,4.70i,4.74i$.

At both gain--loss strengths, the spectral magnitudes obtained with the
three cutoffs nearly overlap. Taking $\varepsilon=10^{-10}$ as the
reference, the relative complex errors over the three sampled energies
are $\sim 10^{-4}$ and $\sim 10^{-5}$ for
$\varepsilon=10^{-6}$ and $10^{-8}$, respectively, at $\lambda=1.4$.
The corresponding errors at $\lambda=5$ are $\sim 10^{-4}$ and
$\sim 10^{-6}$. Thus, tightening the cutoff from $10^{-8}$ to
$10^{-10}$ produces only a small change in the sampled peak regions,
supporting the use of $\varepsilon=10^{-8}$ in the other calculations.

The recorded MPS bond dimensions in Fig.~\ref{fig:sm_cutoff}(d) further reveal two distinct truncation
regimes. At $\lambda=1.4$, both the ordinary and derivative vectors reach
the imposed cap $\chi=100$ for all three cutoffs, indicating that the
bond-dimension cap rather than the compression cutoff limits the retained
MPS dimension. At $\lambda=5$, neither sequence reaches the cap. The
largest recorded bond dimensions of the ordinary vectors are
$23$, $40$, and $71$ as the cutoff is tightened from $10^{-6}$ to
$10^{-8}$ and $10^{-10}$, respectively, while the corresponding
derivative-vector dimensions are $22$, $38$, and $63$.
The relative importance of the compression cutoff and the bond-dimension
cap therefore depends on the spectral regime.
This fixed-$\chi$ comparison quantifies the residual cutoff sensitivity
at the selected energies. It is not intended as an estimate of the error
relative to an uncompressed recurrence or as a full-spectrum error bound.
Overall, the results show that the $\varepsilon=10^{-8}$ cutoff introduces
only a minor numerical dependence in the spectral regions examined here.

\section{Numerical implementation and Hamiltonian validation}
\label{sec:sm_resources}

\subsection{Tensor representation, recurrence, and spectral rescaling}

\begin{figure*}[!t]
\centering
\includegraphics[width=\textwidth]{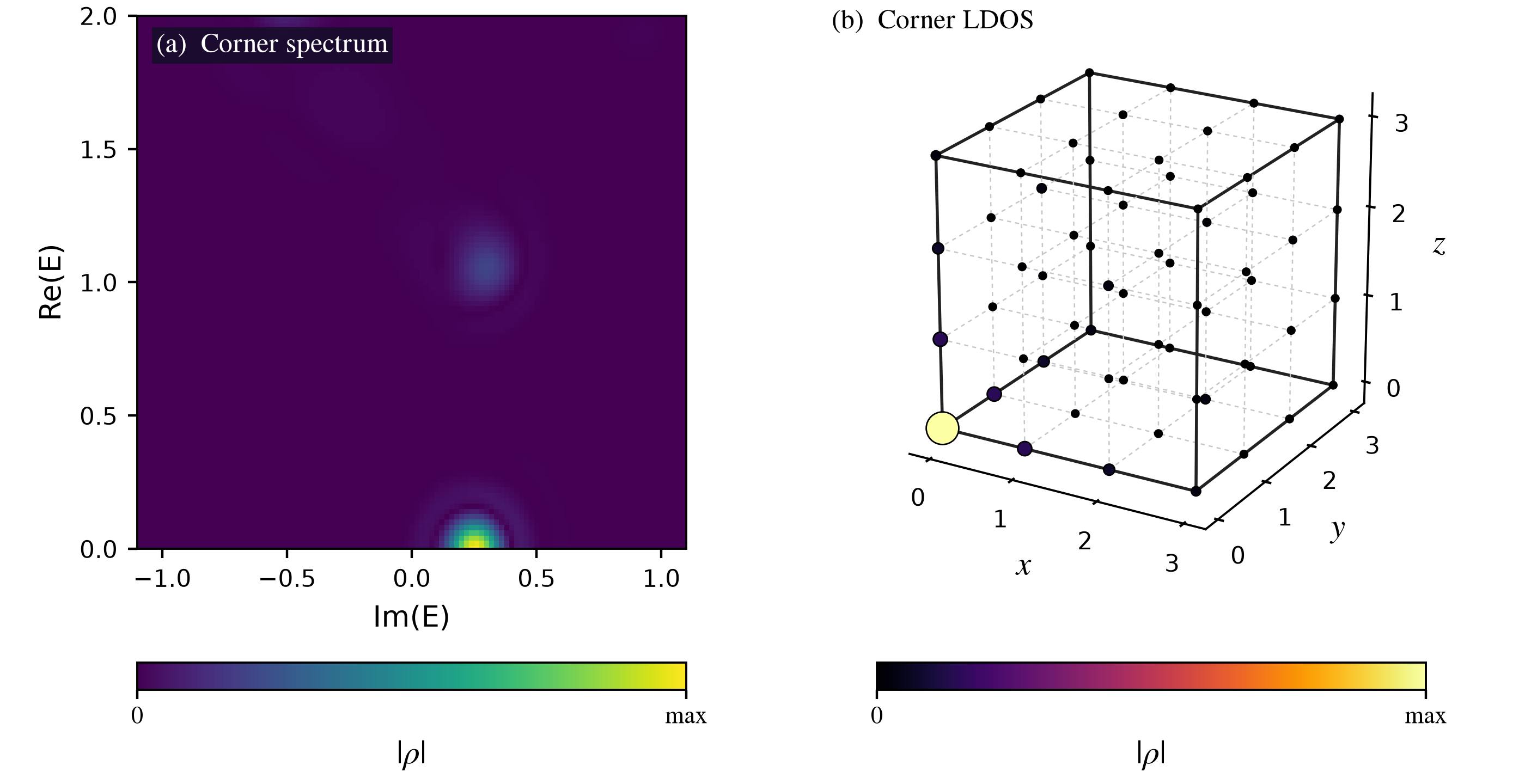}
\caption{Spectral and spatial response of the open $16384^3$ lattice with
$\lambda=1.4$, $t_1=1$, $t_2=1.4$, $N_c=500$, $a=12$,
$\chi=100$ and compression/QTCI tolerances $10^{-8}$.
(a) Corner spectrum on $\mathrm{Re}(E)\in[0,2]$ and
$\mathrm{Im}(E)\in[-1.1,1.1]$, sampled on a $101\times111$ grid
with step $0.02$. (b) The 64 sites $x,y,z=0,1,2,3$ at the sampled
maximum $E=0.26i$. Color and marker area encode $|\rho|$.
Both colorbars use linear physical amplitude scales. No spatial
interpolation or symmetry averaging is applied.}
\label{fig:sm_trillion_spectrum}
\end{figure*}

For a cubic lattice of side length $\ell=2^q$, the $N=\ell^3$ physical
sites are represented by $L=3q=\log_2N$ quantics sites. Coordinate bits
are grouped into $z/y/x$ blocks, with $x$ the fastest real-space
coordinate. Hermitization adds one auxiliary two-level site. The
Hamiltonian MPO bond dimension $D_H$ is distinct from the MPS bond
dimension $\chi$ used in the recurrence. In the periodic-potential
resource calculations below, $D_H=5$, while the Hermitized MPO has
bond dimension $6$. 

The calculations are implemented in Julia 1.10.3 with ITensors.jl
0.9.10 and ITensorMPS.jl 0.3.20~\cite{itensor,ITensorsRepository,ITensorMPSRepository}.
MPO--MPS applications and MPS sums in the recurrence use the
density-matrix compression algorithm with maximum bond dimension $\chi$
and cutoff $10^{-8}$, unless stated otherwise. The Hamiltonian MPO is
constructed with QuanticsTCI.jl 0.7.2 and TensorCrossInterpolation.jl
0.9.17~\cite{NunezFernandez2025learning,QuanticsTCIRepository,TensorCrossInterpolationRepository}.
The public code evaluates Eq.~\eqref{eq:LDOS_MPS} with the prefactor
$2/\pi^2$ of Ref.~\cite{PhysRevLett.130.100401}. The absolute values
of $|\rho|$ shown in this work include the resulting factor $\pi$, so
that the spectral weight of each eigenvalue integrates to one.
Main-text Figs.~\ref{fig:3} and \ref{fig:4} were computed with the
original implementation of the public code, which forms the
kernel-weighted sum of the derivative MPS before the final overlap.
In this implementation, the MPS additions within the recurrence use the
cutoff $10^{-8}$, whereas the MPO--MPS applications and the final
kernel-weighted MPS sum use the bond cap $\chi$ with the
package-default cutoff of $10^{-8}$.
All
appendix calculations use the scalar-moment accumulation of
Eq.~\eqref{eq:LDOS_MPS}. At the overlapping corner energies,
$E=0.25i$ for $\lambda=1.4$ and $E=4.60i$--$4.75i$ for $\lambda=5$,
the two implementations agree to relative differences of order
$10^{-5}$ or smaller.

All KPM spectra use $a=12$, fixed across each sampled energy window.
For the present short-range tight-binding Hamiltonian, the local hopping
amplitudes, onsite terms, and coordination number remain fixed as the
system size is increased, implying an $\ell$-independent upper bound on the
operator norm of $\omega I-H$. The required rescaling factor therefore
does not need to grow with the lattice size.
As an explicit finite-size check, we evaluate the largest singular value
of $\omega I-H$, which equals the largest positive eigenvalue of the
corresponding unscaled Hermitized matrix, on the $16^3$ lattice at
$\lambda=5$. At $\omega=5i$, the Hermitized spectrum extends from
$-11.081$ to $11.081$, giving a rescaled spectral radius of $0.9234$.
We also check all four vertices of the full energy window
$\operatorname{Re}(\omega)\in[-1,1]$ and
$\operatorname{Im}(\omega)\in[0,5]$ shown in Fig.~\ref{fig:4}.
The largest singular value is $11.2102$, attained at
$\omega=\pm1+5i$, corresponding to a rescaled spectral radius of
$0.9342$. Since the operator norm of $\omega I-H$ is convex in the real
and imaginary parts of $\omega$, these vertex values bound the entire
rectangular window. Thus, $a=12$ keeps the rescaled spectrum inside
$(-1,1)$ throughout this window for the tested $16^3$ system.
More generally, once the Hermitized operator has been constructed as an
MPO, its spectral bound could be estimated directly in the
tensor-network representation, for example, with the density-matrix renormalization group (DMRG).
This provides a
practical way of determining or verifying the required rescaling bound
for larger or more complicated systems when an explicit matrix
calculation is unavailable.

\subsection{Energy grids and spatial probes}
\label{sec:sm_sampling}

For Fig.~\ref{fig:3}(a), the corner spectrum at $\lambda=1.4$ is sampled over
$\operatorname{Re}(E)\in[0,1]$ and $\operatorname{Im}(E)\in[0,2]$,
with spacing $0.05$ in both directions. This gives a $21\times41$
grid of 861 distinct complex energies. Figure~\ref{fig:3}(b) shows
the 41-point $\operatorname{Re}(E)=0$ cut, and Fig.~\ref{fig:3}(c) samples
the 64 sites of the outer $4\times4\times4$ corner cube at $E=0.25i$.

For Fig.~\ref{fig:4}(a), at $\lambda=5$, the specified grid covers
$\operatorname{Re}(E)\in[0,1]$ and $\operatorname{Im}(E)\in[0,5]$
with the same spacing $0.05$, giving $21\times101=2121$ distinct
complex energies per symmetry-inequivalent vertex class. The
eight-vertex corner average uses four such classes with multiplicities
$1,3,3,1$. In both spectral maps, the negative-real-energy half is
obtained by symmetry reflection and does not add independently
calculated energy points. The $\operatorname{Re}(E)=0$
line cut in Fig.~\ref{fig:4}(b) contains 101 energies,
$\operatorname{Im}(E)=0,0.05,\ldots,5.00$. These coordinates
are already part of the two-dimensional grid. Figures~\ref{fig:4}(c)--\ref{fig:4}(f) each sample 64 corner-cube sites at a fixed energy,
$E=4.70i,4.50i,4.25i,4.00i$, respectively.

The matched $16^3$
periodic-model spectral benchmark uses a $101\times121$ grid, or
12,221 energies. The $16384^3$ spectrum uses a $101\times111$ grid,
or 11,211 energies. The domain-wall comparison uses a $451\times161$
grid, or 72,611 energies for each of the two probes. The corresponding
energy ranges and spacings are specified in the figure captions.
The resource-scaling study samples three energies at each of six sizes,
giving 18 single-point measurements under the same hardware and software
conditions, rather than timing each complete spectral grid.

\subsection{Hamiltonian MPO validation}

\begin{figure}[!t]
\centering
\includegraphics[width=\linewidth]{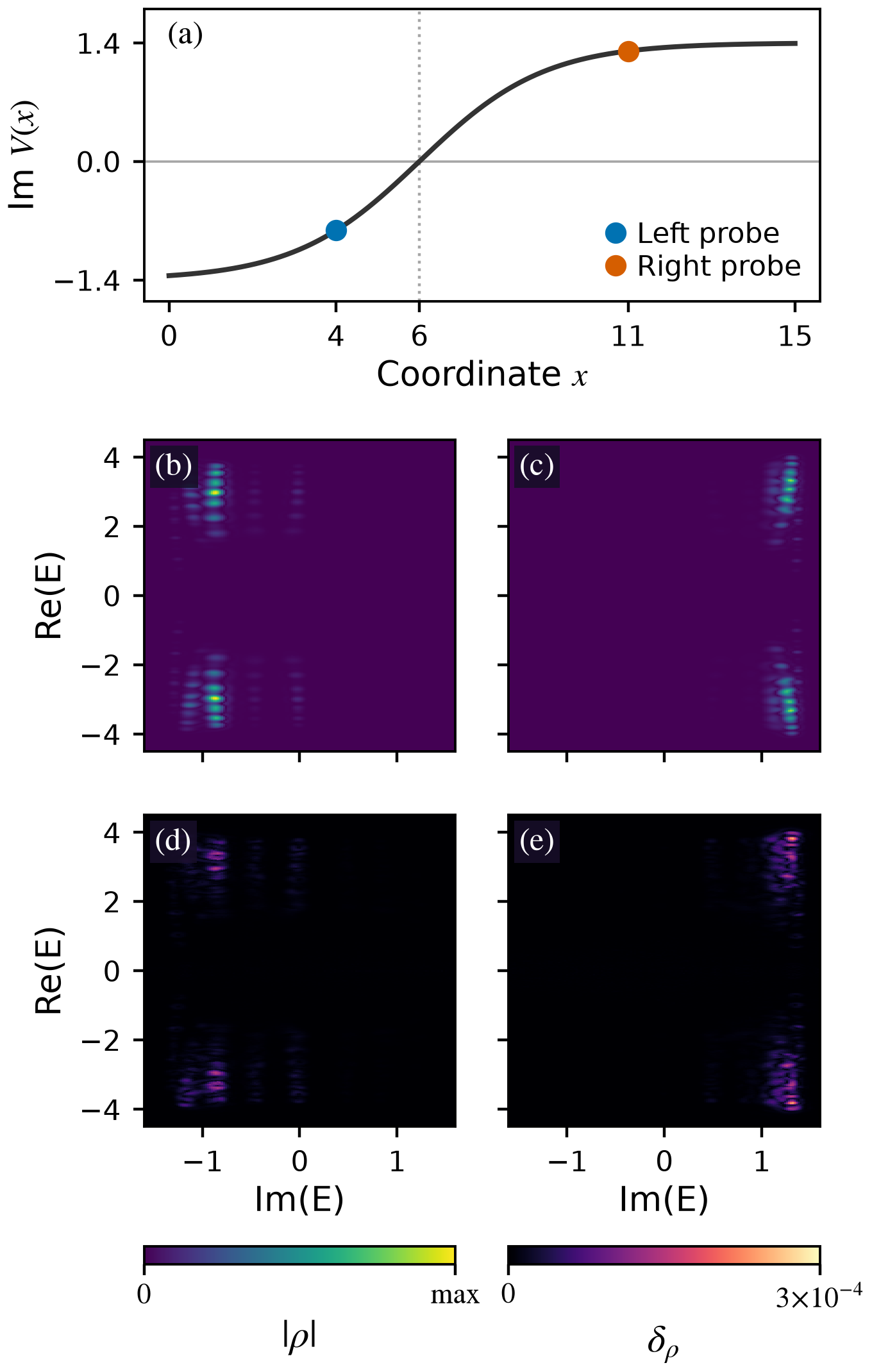}
\caption{TN calculation and sparse-KPM validation for the smooth
domain wall on the open $16^3$ lattice.
(a) Imaginary onsite potential of Eq.~\eqref{eq:sm_domainwall}, with
$\gamma=1.4$, $x_0=6$, and $R=3$. Blue and orange markers identify the
left probe $(4,7,7)$ and right probe $(11,7,7)$.
(b,c) Corresponding TN local spectra, respectively, sharing one linear
amplitude scale labeled from $0$ to max.
(d,e) Corresponding complex-data errors
$\delta_\rho=|\rho_{\mathrm{TN}}-\rho_{\mathrm{sparse}}|/
\max_E|\rho_{\mathrm{sparse}}|$, with a separate denominator for each
probe and one common error scale. Both methods use $N_c=500$ and $a=12$.
TN uses $\chi=100$ and compression/QTCI tolerances $10^{-8}$.
The energy grid spans
$\mathrm{Re}(E)\in[-4.5,4.5]$ and $\mathrm{Im}(E)\in[-1.6,1.6]$,
with step $0.02$ in both directions.}
\label{fig:sm_domain_tn}
\end{figure}

To independently validate the Hamiltonian MPO construction, we assemble the corresponding sparse Hamiltonian and contract the complete production
MPO to a $4096\times4096$ matrix on the open $16^3$ lattice for
$\lambda=2$ and $5$. The comparison includes all matrix elements,
including those that vanish in the exact sparse representation, and uses
the same $z/y/x$ quantics ordering and $x$-fastest real-space indexing.
The QTCI construction tolerance is $10^{-8}$. This validation is performed
before Hermitization and is therefore independent of the subsequent MPS
recurrence.
The relative Frobenius error is defined as
$
\Delta_{\text{MPO}} = 
\frac{\|H_{\rm MPO}-H_{\rm sparse}\|_F}
     {\|H_{\rm sparse}\|_F}
$.
We obtain $\Delta_{\text{MPO}}\sim 10^{-14}$ for both $\lambda=2$ and $5$, with maximum absolute
entrywise errors $\sim 10^{-13}$ and a maximum MPO bond dimension of $D_H=5$.
These results confirm that the QTCI-constructed MPO reproduces the
independently assembled Hamiltonian to near machine precision on the
tested lattice.

\subsection{Bond-dimension growth during the recurrence}

Figure~\ref{fig:sm_bond_growth} tracks the largest bond dimensions of
the ordinary and derivative vectors, $|t_n\rangle$ and $|d_n\rangle$,
as the Chebyshev recurrence proceeds through the recorded odd orders.
This provides a direct measure of the complexity of the MPS objects
propagated during the calculation, complementary to the previous convergence
analysis.

For the $16^3$ lattice, the retained bond dimensions remain below the
imposed cap and reach a maximum value of $64$. In contrast, both the
$1024^3$ and $16384^3$ calculations reach the imposed cap
$\chi=100$ during the recurrence. The large-system calculations are
therefore explicitly bond-dimension limited. Once the cap is reached,
further entanglement growth is absorbed by the MPS truncation rather
than by increasing the size of the propagated tensors. The recorded
bond dimensions should consequently be interpreted as the dimensions
retained by the numerical scheme, rather than as estimates of the
untruncated bond dimensions required by the exact recurrence.

The numerical effect of this truncation is assessed independently by
the $\chi$-convergence calculations in
Fig.~\ref{fig:sm_chi}, where increasing the cap from $\chi=100$ to
$150$ produces only a small change in the sampled spectral region.
Figure~\ref{fig:sm_bond_growth}, by contrast, characterizes the
computational complexity of the propagated MPS and shows that the
large-scale calculations operate in a fixed-$\chi$ truncation regime.
This bounded tensor size is a key ingredient underlying the favorable
resource scaling discussed below.

\section{Computational reach and spatially varying potentials}
\label{sec:sm_capabilities}

The following calculations examine two complementary capabilities of the
quantics--NHKPM framework. These capabilities involve accessing exceptionally large real-space
systems through compressed propagation, and constructing Hamiltonian
MPOs directly from spatially varying coefficients using QTCI. The
resource-scaling analysis and trillion-site calculation quantify the
first capability, while the smooth domain-wall example tests the second
on a system for which the resulting local spectra can be independently
verified using sparse-matrix KPM. Together, these calculations assess
both the accessible system size and the flexibility of the real-space
Hamiltonian construction.

\subsection{Scaling with physical system size}

Figure~\ref{fig:sm_scaling} reports the measured recurrence time
and peak process resident memory at
$E=0.20i,0.25i,0.30i$ for
$\ell=16,64,256,1024,4096,16384$, at fixed
$\lambda=1.4$, $N_c=500$, and $\chi=100$. Over this range, the number
of physical sites increases by a factor of $2^{30}$, whereas the quantics
chain grows only from 12 to 42 sites before Hermitization. For the largest
system, $\ell=16384$, the recurrence requires approximately 67 minutes per
energy point, and the measured peak resident set size (RSS) of the Julia
process is $1.92$--$1.96$ GiB across these three sampled energies.
All scaling measurements use one CPU thread on Intel Xeon E5-2680 v3
processors at 2.50 GHz and the same software environment. The reported timing
includes the KPM recurrence and first-call compilation within this stage,
but excludes Hamiltonian construction and Hermitization. The RSS is a
process-level measurement, including runtime and temporary allocations.
It is not the storage size of the MPS tensors alone. These measurements
characterize the TN implementation. We have not performed a matched
sparse-KPM timing benchmark at the largest sizes, so no runtime speedup
over sparse KPM is inferred from the storage comparison.

\subsection{Trillion-site spectral and spatial response}

We extend the calculation to
\[
N=16384^3=2^{42}\simeq4.4\times10^{12}
\]
physical sites, more than four trillion sites. Relative to the
original $1024^3=2^{30}$ calculation, this requires only 12 additional
quantics sites before Hermitization. At 16 bytes per double-precision
complex entry, a single explicit physical-space vector requires
$16N$ bytes, or $64$ TiB. A single vector in the doubled Hermitized
space requires $32N$ bytes, or $128$ TiB, with $1\,\mathrm{TiB}=2^{40}$
bytes. These are analytical storage lower bounds for an implementation
that retains the corresponding full vector, before additional recurrence
vectors, Hamiltonian storage, and workspace are included. They refer to
total vector storage, which may be distributed across nodes, and are not
measured sparse-KPM process-memory requirements.
The TN resource measurements at this system size give a peak process
RSS of $1.92$--$1.96$ GiB for the three energies and truncation settings
specified above. This comparison quantifies the memory advantage of
the compressed representation for this model without equating a
vector-storage lower bound with a full-process measurement or a runtime
benchmark.
Figure~\ref{fig:sm_trillion_spectrum} shows the computed spectral function for the trillion-site system.
The sampled complex-energy window contains 11,211
energy points, and the largest corner response occurs at $E=0.26i$ on
the $0.02$ energy grid. We then independently evaluate the 64 sites in
the outer $4\times4\times4$ corner cube at this energy. The response is
strongly localized near the exterior corner,
as observed in smaller systems.

\subsection{QTCI construction for a smooth gain--loss domain wall}
\label{sec:sm_domainwall}

In addition to the accessible system size, we examine the applicability
of the method to spatially nonuniform Hamiltonians. We use a smooth
gain--loss domain wall to demonstrate the direct QTCI construction of a
nonperiodic potential from function evaluations and validate the
resulting spectra against sparse KPM.

Specifically, we replace the periodic gain--loss term
$i\lambda V(\mathbf r)$ of Eq.~\eqref{eq:gain_loss} by the onsite potential
\begin{equation}
V(x)=i\gamma\tanh\!\left(\frac{x-x_0}{R}\right),
\label{eq:sm_domainwall}
\end{equation}
on an open $16^3$ lattice, while keeping the hopping terms unchanged.
We use $\gamma=1.4$, $x_0=6$, and $R=3$, and probe the sites
$(4,7,7)$ and $(11,7,7)$ on opposite sides of the domain wall.
The function is sampled at the discrete lattice coordinates and
tensorized by QTCI to construct the diagonal potential MPO. The moderate
system size allows the resulting spectra to be independently validated
against a directly assembled sparse Hamiltonian.
Figure~\ref{fig:sm_domain_tn}(b,c) shows distinct local spectral distributions
on the two sides of the domain wall. The dominant sampled peaks occur near
$E=\pm2.98-0.86i$ for the left probe and
$E=\pm3.32+1.30i$ for the right probe, reflecting the different local
gain--loss environments. Their imaginary parts are close to the local
values $\mathrm{Im}\,V=-0.82$ and $1.31$ at the two probes. The QTCI-constructed gain--loss potential MPO
has bond dimension $4$, while the complete Hamiltonian MPO and its
Hermitized form both have bond dimension $7$.

As a direct check of the tensorized spatial coefficient, the largest
discrepancy between the sampled potential values and those reconstructed
from the MPO is $\sim 10^{-16}$. This verifies the QTCI construction
of the nonuniform potential independently of the subsequent KPM
recurrence. Thus, the spatially varying coefficient can be incorporated
directly from pointwise function evaluations without deriving a dedicated
analytic MPO for the chosen profile.

Panels (d,e) of Fig.~\ref{fig:sm_domain_tn} quantify the differences
between TN and sparse-matrix KPM using the same onsite potential and KPM
parameters. The errors are evaluated from the saved complex data over the
complete grid, including the low-weight background.
The two methods reproduce the same left--right spectral contrast over
the complete complex-energy grid. The relative complex errors over all
72,611 sampled energies per probe are $\sim 10^{-4}$ on both sides. The largest pointwise
differences, normalized by the sparse-KPM peak at each probe, are
also $\sim 10^{-4}$ on both sides.

Together with the direct potential-reconstruction check above, these
results validate both the QTCI construction of the spatially varying
Hamiltonian and its subsequent use in the TN-KPM calculation. Importantly,
the MPO is generated directly from pointwise evaluations of the spatial
coefficient, without requiring a model-specific analytic MPO
representation. This provides a practical route to treating more general
nonperiodic and incommensurate real-space Hamiltonians, complementing the
large-system capability demonstrated in
Figs.~\ref{fig:sm_scaling} and~\ref{fig:sm_trillion_spectrum}.

\FloatBarrier

\bibliography{merged}

\end{document}